\documentclass[aps,pra,floatfix,twocolumn,longbibliography, superscriptaddress]{revtexFORTABLE}

\usepackage{pifont}
\usepackage[T1]{fontenc}
\usepackage[dvipsnames]{xcolor}
\definecolor{bananayellow}{rgb}{1.0, 0.88, 0.21}
\definecolor{amethyst}{rgb}{0.6, 0.4, 0.8}
\definecolor{ao(english)}{rgb}{0.0, 0.5, 0.0}

\usepackage{float}
\usepackage{epsfig}
\usepackage{bm}

\usepackage[matrix,frame,arrow]{xy}

\usepackage[applemac]{inputenc}
\usepackage[T1]{fontenc}
\usepackage{lmodern}
\usepackage[english]{babel}
\usepackage{amsmath}
\usepackage{ae}
\usepackage{units}
\usepackage{amssymb}
\usepackage{color}
\usepackage{graphicx}
\usepackage{bbm}
\usepackage{url}
\usepackage{nomencl}
\usepackage{subfigure}
\usepackage{slashed}
\usepackage{soul}
\usepackage{wrapfig}
\usepackage{amsmath}

 \usepackage{array,multirow}

\usepackage{fixmath}
\usepackage{booktabs}
\usepackage{mathtools}

\usepackage{mleftright}

\newcommand{\bra}[1]{\langle #1 |}
\newcommand{\ket}[1]{| #1 \rangle}  
\newcommand{\braket}[2]{\langle #1 | #2 \rangle}
\newcommand{\ketbra}[2]{\left| #1 \rangle \langle #2 \right|}
\newcommand{\brakket}[3]{\langle #1 | #2 | #3 \rangle}

\newcommand{\expec}[1]{\langle #1 \rangle}

\newcommand{\abssq}[1]{\left| #1 \right|^2}

\newcommand{\figref}[1]{\mbox{Fig.~\ref{#1}}}

\newcommand{\secref}[1]{\mbox{Sec.~\ref{#1}}}

\newcommand{\appref}[1]{\mbox{Appendix~\ref{#1}}}
\renewcommand{\eqref}[1]{\mbox{Eq.~(\ref{#1})}}

\newcommand{\figpanel}[2]{Fig.~\hyperref[#1]{\ref*{#1}(#2)}}
\newcommand{\figpanels}[3]{Fig.~\hyperref[#1]{\ref*{#1}(#2)-(#3)}}
\newcommand{\figpanelNoPrefix}[2]{\hyperref[#1]{\ref*{#1}(#2)}}

\newcommand{\be}{\begin{equation}}
\newcommand{\ee}{\end{equation}}
\newcommand{\bea}{\begin{eqnarray}}
\newcommand{\eea}{\end{eqnarray}}

\usepackage[colorlinks]{hyperref}
\hypersetup{%
	plainpages=true,
	breaklinks=true,
	hypertexnames=false,
	pageanchor=true,
	colorlinks=true,
	linkcolor={blue},
	citecolor={red},
	urlcolor={blue},
	anchorcolor={black}
}

\usepackage[normalem]{ulem}

\begin{document}

\title{Revealing higher-order light and matter energy exchanges using quantum trajectories in ultrastrong coupling}

\author{Vincenzo Macr\`{i}}
\thanks{Equal author contributions}
\email{vincenzo.macri@riken.jp}
\affiliation{Theoretical Quantum Physics Laboratory, RIKEN, Wako-shi, Saitama 351-0198, Japan}

\author{Fabrizio Minganti}
\thanks{Equal author contributions}
\email{fabrizio.minganti@riken.jp}
\affiliation{Theoretical Quantum Physics Laboratory, RIKEN, Wako-shi, Saitama 351-0198, Japan}
\affiliation{Institute of Physics, Ecole Polytechnique F\'ed\'erale de Lausanne (EPFL), CH-1015 Lausanne, Switzerland}

\author{Anton Frisk Kockum}
\affiliation{Department of Microtechnology and Nanoscience, Chalmers University of Technology, 412 96 Gothenburg, Sweden}

\author{Alessandro Ridolfo}
\affiliation{Dipartimento di Fisica e Astronomia, Universit\`{a} di Catania, 95123 Catania, Italy}
\affiliation{INFN Sezione Catania, Catania, Italy}

\author{Salvatore Savasta}
\affiliation{Theoretical Quantum Physics Laboratory, RIKEN, Wako-shi, Saitama 351-0198, Japan}
\affiliation{Dipartimento di Scienze Matematiche e Informatiche, Scienze Fisiche e Scienze della Terra, Universit\`{a} di Messina, I-98166 Messina, Italy}

\author{Franco Nori}
\affiliation{Theoretical Quantum Physics Laboratory, RIKEN, Wako-shi, Saitama 351-0198, Japan}
\affiliation{RIKEN Center for Quantum Computing (RQC), Wakoshi, Saitama 351-0198, Japan}
\affiliation{Physics Department, The University of Michigan, Ann Arbor, Michigan 48109-1040, USA}

\date{\today}

\begin{abstract}

The dynamics of open quantum systems is often modelled using master equations, which describe the expected outcome of an experiment (i.e., the average over many realizations of the same dynamics). 
Quantum trajectories, instead, model the outcome of \textit{ideal} single experiments---the ``clicks''  of a perfect detector due to, e.g., spontaneous emission. The correct description of quantum jumps, which are related to random events characterizing a sudden change in the wave function of an open quantum system, is pivotal to the definition of quantum trajectories.
In this article, we extend the formalism of quantum trajectories to open quantum systems with ultrastrong coupling (USC) between light and matter by properly defining jump operators in this regime. In such systems, exotic higher-order quantum-state- and energy-transfer can take place without conserving the total number of excitations in the system. 
The emitted field of such USC systems bears signatures of these higher-order processes, and significantly differs from similar processes at lower coupling strengths. Notably, the emission statistics must be taken at a single quantum trajectory level, since the signatures of these processes are washed out by the ``averaging'' of a master equation.
We analyze the impact of the chosen unravelling (i.e., how one collects the output field of the system) for the quantum trajectories and show that these effects of the higher-order USC processes can be revealed in experiments by constructing histograms of detected quantum jumps. We illustrate these ideas by analyzing the excitation of two atoms by a single photon~[Garziano et al., \href{https://journals.aps.org/prl/abstract/10.1103/PhysRevLett.117.043601}{Phys.~Rev.~Lett.~\textbf{117}, 043601 (2016)}]. For example, quantum trajectories reveal that keeping track of the quantum jumps from the atoms allow to reconstruct both the oscillations between one photon and two atoms, as well as emerging Rabi oscillations between the two atoms. 
\end{abstract}

\maketitle
\tableofcontents

\section{Introduction}

\subsection{Interacting quantum systems, ultrastrong coupling, and virtual processes}
In a system consisting of two (or more) interacting subsystems, coherent energy transfer can take place between these subsystems. 
If the interaction is small and the subsystems are resonant, a single excitation can be exchanged and the total number of excitations is conserved along the dynamics.
Instead, if the interaction strength is \textit{ultrastrong}~\cite{Kockum2019, Forn-Diaz2019}, i.e.,  comparable to the bare transition frequencies of the individual subsystems,
novel quantum processes can be realized, where the excitation number is not conserved~\cite{Kockum2017, Kockum2019}. 
In this regime, the transition from an initial state $\ket{i}$ to a final state $\ket{f}$, 
characterized by different numbers of excitations, but whose energy is comparable, can take place through a series of \textit{virtual} transitions (intermediate states).
The effective $\ket{i} \to \ket{f}$ process can be described by an effective interaction potential, whose form can be determined by perturbation theory involving a sum over all the possible contributing virtual transitions.

Processes mediated by virtual transitions are common also in open quantum systems, where the energy-conservation condition is relaxed by the inclusion of dissipation. For example, in nonlinear quantum optics~\cite{Boyd2008} and polaritonics \cite{Ciuti2013a}, the $\chi^{(3)}$ interaction is due to the virtual creation of an electron-hole pairs. 
Similarly to the Hamiltonian case, also in open quantum system an effective Hamiltonian can capture an emergent coupling between different states. While the dynamics of the closed system is completely determined by the effective Hamiltonian, in the open-system case, the presence of dissipation can mix different Hamiltonian manifolds and affect the dynamics in nontrivial ways. 

An interesting example of a system with virtual transitions is that of ultrastrong coupling (USC) between light and matter.
While the Hamiltonian processes are characterized by USC, the elecromagnetic field cannot be isolated from the environment, resulting in an open system dynamics.
The USC regime was defined for intersubband polaritons~\cite{Ciuti2005} and experimentally observed in a microcavity-embedded doped GaAs quantum well~\cite{Anappara2009} and in circuit quantum electrodynamics (QED)~\cite{Niemczyk2010}.
After that, USC has been reached in several others experimental platforms, including cavity QED and circuit optomechanics (see Refs.~\cite{Kockum2019,Forn-Diaz2019,Kockum2019a} and references therein). 
Following these experimental developments, interest in USC has blossomed, stimulating many theoretical studies~\cite{Liberato2007a, Ashhab2010, Cao2010, Casanova2010, Braak2011, Ridolfo2012, Stassi2013, DeLiberato2014, Sanchez2014, Garziano2014a, Cirio2016, Macri2016, garziano2017a, DiStefano2017a, Macri2018a, DiStefano2018, DeBernardis2018, DiStefano2019a, Stokes2020, LeBoite2020, Pilar2020, Felicetti2020, Ashida2021}. In particular, 
processes that do not conserve the total number of excitations have attracted considerable attention~\cite{Law2015, Garziano2015, Kockum2017a, Stassi2017a, Macri2018, DiStefano2019}. Among them, the possibility of single photons simultaneously exciting two or more atoms~\cite{Garziano2016, Macri2020,GarzianoSciRep2020}
will be used in this article as an illustrative example. 
This intriguing process arises from the interplay of a complex combination of higher-order virtual processes.

\subsection{Quantum trajectories}

For a weakly coupled Markovian environment, the physics of an open quantum system is described by a Lindblad master equation (LME)~\cite{Carmichael1993, Breuer2002, Haroche2006, Wiseman_BOOK_Quantum}. The state of the system evolving with the LME is captured by the density matrix, which represents the average state of the system over many experiments. The effect of the environment on the system is described via an ensemble of \textit{quantum jumps} acting on the density matrix through the dissipation superoperators.

Although the physics of the system can be encoded by the LME, this theoretical treatment does not allow for an easy description of a single experiment. 
For this purpose, the stochastic evolution of the system's wave function constitutes an efficient alternative to the LME approach
~\cite{Gisin1984, diosi1986, Gardiner1992}. In quantum trajectories, the interaction between the system and its environment is modelled as a set of ideal detectors, which continuously monitor the output field of the system~\cite{Wiseman_BOOK_Quantum}. 
Quantum jumps have been observed in many experimental platforms, ranging from solid-tate physics to superconducting circuits (see, e.g., Refs.~\cite{NagourneyPRL86, SayrinNat11,PeilPRL99,JelezkoAPL02, Minev2019}).
The stochastic evolution of the wave function under such a procedure is known as a \textit{quantum trajectory}~\cite{Dalibard1992, CarmichaelPRL93, Molmer1993}. Since the LME describes the average evolution of the system, it can be obtained by averaging over an infinite number of quantum trajectories.

Even if the LME and quantum-trajectory approaches are equivalent on average, there may exist behaviours witnessed by single quantum trajectories that cannot be directly observed at the LME level because: (i) spontaneous decay processes, induced by the environment, occur randomly and averaging can cancel several features; (ii) there can be rare processes whose visibility is reduced by averaging.
Examples of such processes have been found in bosonic and spin systems, both concerning the states explored by the dynamics and the emergence of different timescales~\cite{Haroche2006, Bartolo2017, Rota2018, Munoz2019B,MingantiPRA2021}. 
\textit{The first goal of this article is the study of how such hidden processes can be used to reveal USC in open quantum systems.}

Experimentally, there exist different ways in which the output field of a cavity can be monitored. Theoretically, this translates into different types of evolution for the quantum trajectories~\cite{Plenio1998}. One such type is a non-Hermitian continuous time evolution interrupted by abrupt changes in the wave function due to quantum jumps. This is the widely used Monte-Carlo-wave-function (MCWF) method~\cite{Molmer1993}. Another type of evolution is continuous stochastic infinitesimal changes of the wave function due to a noise term. This is the quantum-state-diffusion (QSD) method~\cite{Gisin1992, gisin1993, GisinAA1993, Percival2002}. For photons escaping an electromagnetic resonator, the MCWF method describes the ideal photodetection of the output field, while the QSD method describes homodyne measurements.

Furthermore, the access to the emitted field of a USC system allows to reconstruct some of the correlation functions of the system \cite{Fink2018}. In this regard, quantum trajectories allow to predict the presence of USC phenomena by histogramming the statistics of quantum jumps.
While normally this would be a nonessential remark, in USC it is often difficult to reconstruct the presence of higher-order processes, due to both the fragility of these processes with respect to external perturbation (they are higher-order perturbative effects) and to the intrinsic difficulty in measuring the effects of virtual excitations \cite{Lolli2015,Felicetti2015,DeLiberato2009,Cirio2016}.
\textit{The second aim of this article is to show  that an accurate study of quantum trajectories allows to demonstrate the presence of higher-order USC processes.}

\subsection{Outline and original results of this article}

In \secref{sec:Model}, we first present the one-photon--two-atoms system, introduced in Ref.~\cite{Garziano2016}, and explain how higher-order processes allow a single photon to excite two atoms, and vice versa. We provide an effective Hamiltonian for the system we study, and we then show how the formalism of quantum trajectories can be adapted to handle such a USC system. 
Moreover, within this section, we provide analytical results by describing the one-photon--two-atom  and the qubit-qubit processes, where the latter is a second-order sub-process that is part of the main effect.

We use this system as an example to show that \textit{individual quantum trajectories} can clarify the dynamic evolution of interacting quantum systems by revealing \textit{hidden behaviour} that cannot be trivially witnessed by the LME. 
We do it in two cases: In \secref{ResultI}, we consider only local dissipation, while in \secref{single_trajectory_collective_dissipation}, we introduce also a collective dissipation channel for the two qubits \cite{Shammah2018}.

We identify several dynamics stemming from higher-order processes that are revealed by individual quantum trajectories. 
In particular, we show that the quantum jumps back-action induces a dissipative quantum state transfer between the two qubits, similar but not identical to what was shown in Ref.~\cite{MingantiPRA2021}.

In addition,  we show how higher-order processes can be identified also by constructing histograms of detection events.
This is a viable experimental technique, where photodetection from multiple experiments allows to reconstruct the correlation functions. 
Finally, we conclude and give an outlook for future work in \secref{sec:ConclusionOutlook}

In the appendices, we provide a detailed derivation of the effective Hamiltonian for the system we study, a comparison between this effective Hamiltonian and the full system Hamiltonian, and a more detailed analysis of all processes involved. Moreover, we analyze the quantum trajectories that arise when the system output is detected by a homodyne measurement instead of photodetection, demonstrating the importance of the unravelling protocol. We conclude the appendices showing a comparison between the LME and MCWF approaches for obtaining the averaged system dynamics.

Beyond the interest in interacting quantum systems, this article provides a new way to probe the presence of USC effects in a light-matter system, a task which normally is challenging since one cannot directly access the virtual photons populating the dressed states of the system. 

\section{Model and mathematical tools}
\label{sec:Model}

The system studied in Ref.~\cite{Garziano2016} is composed of two subsystems: (i) Two qubits noninteracting with each other; (ii) A single cavity mode. The subsystems are ultrastrongly coupled, and their Hamiltonian is ($\hbar = 1$ throughout this article)
\be
\label{Hamiltonian}
\begin{split}
\hat H &= \omega_c \hat a^\dag \hat a + \frac{1}{2} \sum_i^2  \omega_q^{(i)} \hat \sigma_z^{(i)} \\
&+ g \mleft( \hat a + \hat a^\dag \mright) \sum_i^2 \mleft[ \hat \sigma_x^{(i)} \cos \theta + \hat \sigma_z^{(i)} \sin \theta \mright] \, ,
\end{split}
\ee
where $\hat a^\dag$ ($\hat a$) is the creation (annihilation) operator for the photons in the cavity mode, $\hat \sigma_z^{(i)}$ and $\hat \sigma_x^{(i)}$ are the Pauli operators for the $i$th qubit, and $g$ is the coupling rate of each qubit to the cavity mode. We indicate with  $\ket{n, g, e}$ the state with $n$ photons in the cavity,  qubit 1 in the ground state, and qubit 2 in the excited state.
The Hamiltonian in \eqref{Hamiltonian} is the sum of two elements: a non-interacting part (the first two terms), which describes the bare energy of the subsystems, and the last term, which describes the USC light-matter interaction. Notably, the interaction contains the counter-rotating terms $\sigma_+^{(i)} \hat a^\dag$ ($\sigma_-^{(i)} \hat a$), which create (destroy) two excitations, and $\sigma_z^{(i)} \hat a^\dag$ ($\sigma_z^{(i)} \hat a$), which create (destroy) one excitation. The latter term in \eqref{Hamiltonian} breaks the parity symmetry, and can be realized in superconducting circuits~\cite{Niemczyk2010}.

As shown in Ref.~\cite{Garziano2016}, at the resonance condition $\omega_c \simeq \omega_q^{(1)} + \omega_q^{(2)}$,
the counter-rotating terms enable virtual transitions, allowing the system to oscillate between the two bare states $\ket{1, g, g}$ and $\ket{0, e, e}$, i.e., a single photon can excite both qubits.

\subsection{Effective system Hamiltonian}
\label{sec:EffectiveHamiltonian}

To observe the one-photon--two-atoms process one must avoid the Rabi oscillations between a single qubit and the photonic mode.
As such, the cavity-qubit detuning in \eqref{Hamiltonian} is large compared to the coupling strength: $g\ll (\omega_c - \omega_q^{(i)})$. 

For interacting quantum systems that are strongly detuned, an effective Hamiltonian can be derived  using the generalized James' effective Hamiltonian method~\cite{Shao2017}. 
To apply this method to \eqref{Hamiltonian}, we assume that the bare transition frequencies are close to the resonance condition $\omega_c \simeq \omega_q^{(1)}+ \omega_q^{(2)} = 2 \omega_0$.
With this notation, we indicate that the qubits and cavity have been finely tuned to take into account effective energy shifts induced by the interaction ``dressing" the bare states [see also the discussion in Appendix~\ref{EffHamiltonian}]. 
Thus, considering processes up to third order in the interaction, and neglecting  dressing energy shifts which have been reabsorbed by an appropriate choice of the coefficients, the effective Hamiltonian reads
\be \label{Heff0} 
\hat H_{\rm eff} = \hat H_{\rm eff}^{(2)} + H_{\rm eff}^{(3)} \,.
\ee

By defining the qubit detuning $2 {{{\rm{\Delta}}}} = \omega_q^{(1)} - \omega_q^{(2)}$ (such that $\omega_q^{(1)}=\omega_0 + {{\rm{\Delta}}}$ and $\omega_q^{(2)} = \omega_0 - {{\rm{\Delta}}}$),
we distinguish two regimes of work for the effective Hamiltonian $\hat H_{\rm eff}$: 

(i) Identical qubits (${{\rm{\Delta}}} =0$ and same dissipation rates); 

(ii) Non-identical qubits (${{\rm{\Delta}}} \neq0$ and/or different dissipation rates).

A detailed derivation is provided in \appref{EffHamiltonian}, and in \appref{comparison} we show the excellent agreement between the full model and the effective Hamiltonian near the resonance $\omega_c \simeq \omega_q^{(1)} + \omega_q^{(2)}$ and for small enough ${\rm{\Delta}}$.

\subsubsection{Identical qubits}
\label{sec:identical_qubits}

If ${\rm{\Delta}}=0$, we have
\begin{subequations}

\be \label{Heff_second_oreder}
\hat H_{\rm eff}^{(2)} ={\rm{\Omega}}_{\rm eff}^{(2)} \mleft(\hat \sigma_-^{(1)} \hat \sigma_+^{(2)} +  \hat \sigma_+^{(1)} \hat \sigma_-^{(2)} \mright)\, ,
\ee
\be \label{Heff_third_oreder} 
\hat H_{\rm eff}^{(3)} ={\rm{\Omega}}_{\rm eff}^{(3)}\mleft(\hat a \hat \sigma_+^{(1)} \hat \sigma_+^{(2)} + \hat a^\dag \hat \sigma_-^{(1)} \hat \sigma_-^{(2)} \mright)\, .
\ee
\end{subequations}
The second- and third-order effective Hamiltonains $\hat H_{\rm eff}^{(2, \, 3)}$ represent  second- and third-order perturbative couplings, with effective interactions 
\begin{subequations}
\be\label{Omega_2_identical}
{\rm{\Omega}}_{\rm eff}^{(2)}=- \frac{4  g^2 \cos^2\theta}{3 \omega_0}\;,
\ee
\be\label{Omega_3_identical}
{\rm{\Omega}}_{\rm eff}^{(3)}=-\frac{8 g^3 \cos^2\theta \sin \theta}{3 \omega_0^2}\;.
\ee
\end{subequations}

$\hat H_{\rm eff}^{(2)}$ in \eqref{Heff_second_oreder} is an effective coherent resonant coupling which describes oscillations between the states $\ket{0, e, g}$ and $\ket{0, g, e}$. The coupling ${\rm{\Omega}}_{\rm eff}^{(2)}$ is thus relevant only when ${\rm{\Delta}} \ll {\rm{\Omega}}_{\rm eff}^{(2)}$. As we numerically show in \secref{ResultI} and analytically discuss in \appref{AnalyticalI}, $\hat H_{\rm eff}^{(2)}$ plays an important role when, during the system evolution, one of the two qubits excitations is lost into the environment. 

The third-order effective Hamiltonian in \eqref{Heff_third_oreder} is the one responsible for the one-photon--two-atoms process. Indeed, the term $\hat a \hat \sigma_{+}^{1} \hat \sigma_{+}^{2}$ ($\hat a^{\dag} \hat \sigma_{-}^{1} \hat \sigma_{-}^{2}$) destroys (creates) a photon and simultaneously creates (destroys) two qubit-excitations.
As such,  the states $\ket{1, g, g}$ and $\ket{0, e, e}$ are connected with the effective resonant coupling rate ${\rm{\Omega}}_{\rm eff}^{(3)}$.

\subsubsection{Non-identical qubits}
\label{sec:non_identical_qubits}

If $0<{\rm{\Omega}}_{\rm eff}^{(2)} \ll {\rm{\Delta}}$, the second-order effective interaction $\hat H_{\rm eff}^{(2)}$ in \eqref{Heff0} can be neglected, applying the rotating-wave approximation (RWA). 
However, the third-order effective Hamiltonian can still couple $\ket{1, g, g}$ and $\ket{0, e, e}$ when the resonance condition $\omega_c \simeq \omega_q^{(1)}+ \omega_q^{(2)} = 2 \omega_0$ is satisfied. 
In this case, $H_{\rm eff}= \hat H_{\rm eff}^{(3)}$, where $\hat H_{\rm eff}^{(3)}$ is the one in \eqref{Heff_third_oreder}, but
the coupling rate now is
\bea \label{geff} 
{\rm{\Omega}}_{\rm eff}^{(3)} = - \frac{8 g^3 \cos^2\theta \sin \theta \mleft( 3 \omega_0^2 + {\rm{\Delta}}^2 \mright)}{\mleft( \omega_0^2 - {\rm{\Delta}}^2 \mright) \mleft(9 \omega_0^2 - {\rm{\Delta}}^2 \mright)} \, .
\eea
Notice that the case ${\rm{\Delta}}=0$ can be trivially obtained from \eqref{geff}.

\subsection{Quantum jump operators in the USC regime}
\label{subsection2}

Having derived the effective Hamiltonians, we need to correctly introduce the action of the environment.
Any LME contains a Hamiltonian part, describing a coherent unitary evolution, and a series of dissipators $\mathcal{D}[\hat{O}_m]$ such that
\be
\partial_t \hat{\varrho}= - i [\hat{H}, \hat{\varrho}] +\sum_{m} \gamma_m \mathcal{D}[\hat{O}_m] \hat{\varrho},
\ee
where $\gamma_m$ is the dissipation rate of the operator $\hat{O}_m$ and
\be
\mathcal{D}[\hat{O}_m] \hat{\varrho}=  \hat{O}_m \hat{\varrho} \hat{O}_m^\dagger -\frac{\hat{O}_m^\dagger \hat{O}_m\hat{\varrho} + \hat{\varrho} \hat{O}_m^\dagger \hat{O}_m}{2}.
\ee

\subsubsection{Dressed jump operators}

When dealing with the light-matter coupling, the spontaneous emission in the LME must be modified to take into account the presence of virtual excitations~\cite{LeBoite2020}. 
A general approach to do that was developed in Ref.~\cite{Breuer2002} and has been the workhorse of various other studies of USC dissipative systems~\cite{Beaudoin2011, Ridolfo2012, Bamba2012, Bamba2014, Hu2015, Bamba2016, Settineri2018}.
Every field, coupling the system with the environment, can be expressed as $\hat S_m = \hat s_m+ \hat s_m^\dag$.
When the coupling is not too strong, e.g., a Jaynes--Cummings (JC) model, there are no virtual excitations. Thus, the overall effect of $\hat S_m$ is only to eject excitations into the environment, i.e., $\hat{O}_m =\{\hat{a},\, \hat{\sigma}_{-}\}$.

Instead, in the USC regime a correct treatment of input-output, dissipation, and correlation functions requires that the coupling with the environment does not induce transitions increasing the energy of the system for spontaneous emission.
A physically consistent approach consists of separating each operator $\hat S_m$
into its positive $\hat S_m^+ = \sum_{j, k > j} \brakket{j}{\hat S_m}{k} \ketbra{j}{k}$ and negative $\hat S_m^- = (\hat S_m^+)^\dag$ frequency components.
Those are expanded in terms of the eigenstates $\{|j \rangle,|k\rangle \}$ of the total system Hamiltonian, and $ k > j$ indicates that the energy of $\ket{k}$ is larger than that of $\ket{j}$. This properly defines the jump operators for any arbitrary LME as $\mathcal{D}[\hat{S}^{+}]$, which by construction acts like an excitation annihilation operator.
In this \textit{dressed picture}, the quantum jumps are between the dressed states (the eigenstates) of the system Hamiltonian which, in USC, contain contributions from bare states with an arbitrary number of excitations~\cite{Kockum2019}.

Physically speaking, this procedure amounts to distinguishing between the bare and dressed excitations, i.e., those excitations which cannot be detected versus those which can. 
Not satisfying these conditions leads to the prediction of non-physical behaviours, such as
a continuous emission of photons from the system ground state of an undriven USC system~\cite{Ciuti2006a}. 
As such, when we compute $\expec{\hat{S}_m^{-}\hat{S}_m^{+}}$ we are describing the expected values of the dressed excitations inside the system, which can be emitted into the environment.

\subsubsection{Quantum trajectories in USC}

Having obtained a well-defined  LME, we can now properly introduce the MCWF.
Following Refs.~\cite{Dalibard1992,Molmer1993}, we introduce the non-Hermitian Hamiltonian 
\be \label{non_Hermitian_H} 
\mathcal{\hat H} = \hat H - \frac{i}{2} \sum_m \gamma_m \, \hat S_m^- \hat S_m^+ \, ,
\ee
describing the effect of the environment between two quantum jumps. 
Here, $\hat{H}$ represents the Hamiltonian part of the dynamics, and one can either use the full or the effective Hamiltonian (for the right value of $\Delta$).
The evolution of a quantum trajectory is thus dictated by a non-Hermitian evolution via $\mathcal{\hat H}$ interrupted by random quantum jumps.

The algorithm to obtain such a dynamics reads:

\textbullet \,\, $\ket{\psi (t)}$ is the normalized wave function at the initial time $t$.

\textbullet \, The probability that a quantum jump occurs through the $m$th dissipative channel in a small amount of time $dt$ is
\be
\delta p_m(t) =dt \gamma_m \, \braket{\psi ( t)|\hat S_m^- \hat S_m^+}{\psi (t)},
\ee
such that $\delta p_m(t)\ll 1$.

\textbullet \,\, One randomly generates a real number $\varepsilon \in [0,1]$. 

\textbullet \,\, If $\sum_m \delta p_m(t)<\varepsilon$, no quantum jump occurs, and the system evolves as
\be \label{evolution} 
\ket{\psi (t+ dt)} = \exp \mleft( - i \mathcal{\hat H} dt \mright) =  \mathbbm{1}- i dt \mathcal{\hat H} \ket{\psi (t)} + \mathcal{O}(dt^2) \, .
\ee

\textbullet  \,\, Otherwise, if $\sum_m \delta p_m(t)>\varepsilon$, a quantum jump occurs. To decide which channel dissipates, a second random number $\varepsilon'$ is generated, and each quantum jump is selected with probability $\delta p_m(t)/(\sum_n \delta p_n(t))$.
The wave function then becomes
\be \label{jump} 
\ket{\psi (t + dt)} =  \hat S_m^+ \ket{\psi (t)}
\ee

\textbullet \,\, At this point, independently of whether a quantum jump took place, the wave function $\ket{\psi (t + dt)}$ is renormalized and used for the next step of the time evolution.
\vspace{0.2cm}

Any quantum jump corresponds to the projection of the wave function associated with a generalized measurement process (wave-function collapse through a positive operator-valued measure) \cite{Wiseman_BOOK_Quantum}. 
Although the results of MCWF recovers those of LME by averaging over an infinite number of trajectories, noise effects determine the convergence rate. A discussion on this point is provided in \appref{App:comparison_dynamics}, where we compare the dynamics using both the LME and MCWF approaches for the system under consideration.

\subsection{Analytical results for the time evolution in the general case}

Now  we want to show how, by analyzing a single quantum trajectory, we can analytically describe the phenomena which are taking place and, via the detection of the quantum jumps, reconstruct the one-photon--two-atoms process and the higher-order sub-processes (which are part of the main effect) that are taking place. A more detailed discussion of this analysis can be found in \appref{AnalyticalI}.

We consider four quantum jump operators: 
\be \label{Jump_operators}
\begin{split}
 &\mathcal{D}\left[\sqrt{\kappa} \hat X^+\right], \quad \mathcal{D}\left[\sqrt{\gamma_{1,2}} \hat C_{1,2}^+\right], \\
 & \quad \mathcal{D}\left[\sqrt{\frac{\gamma_C}{2}} (\hat C_1^+ + \hat C_2^+) \right].
\end{split}
\ee
They represent the cavity loss, local qubit de-excitation, and collective qubit emission through a common bath, respectively. Here, $\sqrt{\kappa} \hat X^+$ is the dressed operator for the cavity field, and $\sqrt{\gamma_{1,2}} \hat C_{1,2}^+$ are the qubit ones derived from $\hat \sigma_x^{(i)}$. As such, $\kappa$, $\gamma_{1,2}$, and $\gamma_C$ describe the photon decay rate, individual qubit dissipation, and collective qubit dissipation \cite{Shammah2018,MingantiPRA2021}, respectively.

\subsubsection{One-photon--two-atoms}

\label{One-photon--two-atoms}

By projecting the time-evolution operator $\hat U(t) = \exp \mleft( - i \mathcal{\hat H} t \mright)$ onto the two-dimensional subspace $\{ \ket{1, g, g} , \ket{0, e, e} \}$, one describes the one-photon--two-atom process, which takes place independently of ${\rm{\Delta}}$. In this case, the time-evolution operator $\hat U(t) $ will be 
\begin{widetext}
\be 
\begin{split}
\label{evolution_operator} 
\hat U(t)&= e^{- \frac{1}{4} (\kappa + {\rm{\Gamma}}) t}  \mleft\{ \mleft[ \cos (\eta t / 4) - \frac{\kappa - {\rm{\Gamma}}}{\eta} \sin (\eta t / 4) \mright] \ketbra{1, g, g}{1, g, g} \mright. \\
& \quad - \frac{4 i {\rm{\Omega}}_{\rm eff}^{(3)}}{\eta} \sin (\eta t / 4) \bigg [ \ketbra{1, g, g}{0, e, e} + \ketbra{0, e, e}{1,g,g} \bigg ] 
 \mleft. +\mleft[ \cos (\eta t / 4) + \frac{\kappa - {\rm{\Gamma}}}{\eta} \sin (\eta t / 4) \mright] \ketbra{0, e, e}{0, e, e} \mright\} \, ,
\end{split}
\ee
where ${\eta = \sqrt{\mleft(4 {\rm{\Omega}}_{\rm eff}^{(3)} \mright)^2 - (\kappa - {\rm{\Gamma}})^2}}$, and ${\rm{\Gamma}} = \gamma_1 + \gamma_2 + \gamma_C$ 
is the sum of the qubit loss rates. 
The time evolution operator  $\hat U(t) $  describes the unnormalized oscillations of the wave function. For $\ket{\psi(t)}$ initialized in $\ket{1, g,g}$, we obtain
\be \label{initial_state} 
\begin{split}
&\ket{\psi (t)} =e^{- \frac{1}{4} (\kappa + {\rm{\Gamma}}) t}\mleft\{ \mleft[ \cos (\eta t / 4) - \frac{\kappa - {\rm{\Gamma}}}{\eta} \sin (\eta t / 4) \mright] \ket{1, g, g}  - \frac{4 i {\rm{\Omega}}_{\rm eff}^{(3)}}{\eta} \sin (\eta t / 4) \ket{0, e, e} \mright\} \, .
\end{split}
\ee
\end{widetext} 

By appropriately renormalizing the wave function we obtain the mean photon number $\expec{\hat X^- \hat X^+}$ and mean excitation numbers of the two qubits $\expec{\hat C_i^- \hat C_i^+}$ ($i = 1, 2$): 
\be \label{oscillate_state_cavity}
\begin{split}
\expec{\hat X^- \hat X^+} &= \frac{\cos^2 (\frac{\eta t }{4}) + \mleft( \frac{\kappa - {\rm{\Gamma}}}{\eta} \mright)^2 \sin^2 (\frac{\eta t }{ 4}) - \frac{\kappa - {\rm{\Gamma}}}{\eta} \sin^2 (\frac{\eta t }{2})}{1 - \frac{\kappa - {\rm{\Gamma}}}{\eta} \sin (\frac{\eta t}{2}) + 2 \mleft( \frac{\kappa - {\rm{\Gamma}}}{\eta} \mright)^2 \sin^2 (\frac{\eta t }{ 4})}, \,  \\
\expec{\hat C_i^- \hat C_i^+} &= \frac{\mleft( \frac{4 {\rm{\Omega}}_{\rm eff}^{(3)}}{\eta} \mright)^2 \sin^2 (\frac{\eta t }{ 4})}{1 - \frac{\kappa - {\rm{\Gamma}}}{\eta} \sin (\frac{\eta t }{ 2}) + 2 \mleft( \frac{\kappa - {\rm{\Gamma}}}{\eta} \mright)^2 \sin^2 (\frac{\eta t}{4})} \, .
\end{split}
\ee
If $\eta$ is real, the system oscillates. Interestingly, if one matches the condition $\kappa = {\rm{\Gamma}}$ the system has a purely sinusoidal behavior, and the process takes place with the same rate as in the purely Hamiltonian  case even if the system is dissipative.
If, instead, $\kappa \neq {\rm{\Gamma}}$, the oscillation occurs with a reduced amplitude and a non-sinusoidal shape.
This is a first remarkable prediction of the open-system case: by tuning the dissipation rate, we can deduce the behaviour of the system by considering how the emission statistics (depending on $\expec{\hat X^- \hat X^+}$ and $\expec{\hat C_i^- \hat C_i^+}$) changes.
\subsubsection{First quantum jump}
Knowing the state at time $t$ allows us to predict through which channel, and with which probability, the system is expected to lose an excitation. As such, analyzing the first quantum jump allows to reconstruct the oscillation parameter $\eta$.
Indeed, by repeating the experiment several times, the probability (of a quantum jump to take place) can be reconstructed, and such a probability must oscillate with the same period as $\ket{\psi (t)}$.

\begin{figure*}
	\centering
	\includegraphics[width=0.9\linewidth]{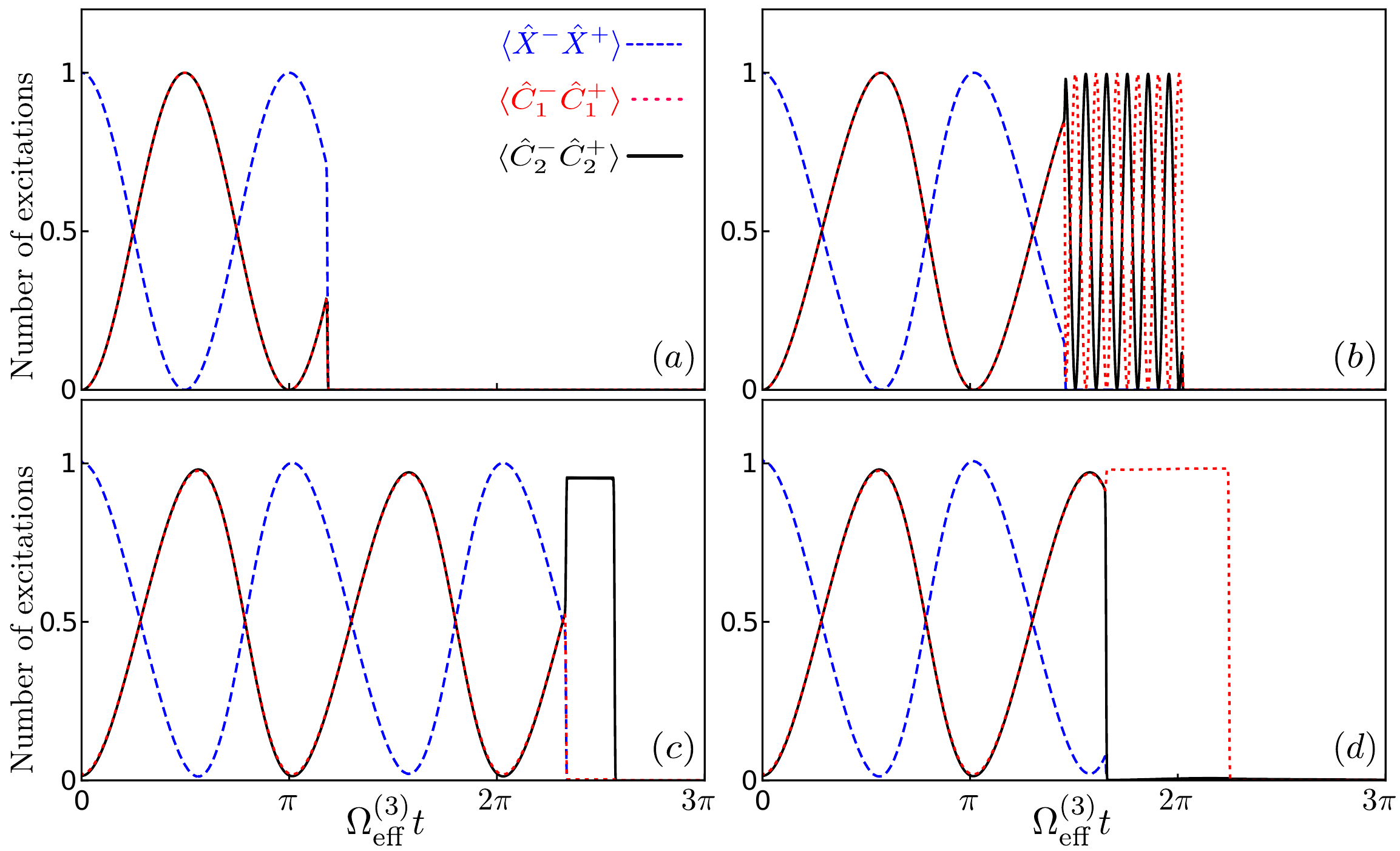}
	\caption{Examples of single quantum trajectories, numerically obtained with the full Hamiltonian in \eqref{Hamiltonian} and the dissipators in  \eqref{Jump_operators} in the absence of collective qubit decay ($\gamma_C=0$). All panels show the time evolution of the mean photon number $\expec{\hat X^- \hat X^+}$ (blue dashed curves) and of the mean excitation numbers of the two qubits $\expec{\hat C_i^- \hat C_i^+}$ ($i = 1, 2$) (red dotted and black solid curves, respectively).
    The system is always initialized in $\ket{1,g,g}$ at the resonant condition $\omega_c \simeq \omega_q^{(1)}+ \omega_q^{(2)} $.  All the panels initially display the oscillation in \eqref{initial_state} until a quantum jump occurs. The panels represent a quantum trajectory where
	(a)  a cavity jump occurs. The detection of an emitted cavity photon projects the wave function onto $\ket{0, g, g}$, where $\expec{\hat X^- \hat X^+} = \expec{\hat C_i^- \hat C_i^+} = 0$;
	(b) Identical qubits case in which a qubit 1 jump occurs, projecting the wave function onto a state with qubit 2 excited. The qubits then start to exchange their excitation between themselves until a second qubit jump takes place, projecting the system onto the state $\ket{0, g, g}$.
	(c) Non-identical qubits case in which a qubit 1 jump occurs. As in (b), the system is projected onto a state where qubit 2 is excited. Being off resonance, this time the qubits do not exchange an excitation and qubit 2 remains excited until a jump projects the system to $\ket{0, g, g}$;
	(d) Non-identical qubits case in which a qubit 2 jump occurs, leading to a dynamics similar to panel (c).
	In all panels, the parameters are $g = 0.1 \omega_0$, $\omega_c \simeq 2 \omega_0$, $\kappa = \gamma_{(1,2)} = 4 \times 10^{-5} \omega_0$, and $\gamma_C = 0$.
	In panels (a)-(b), $\omega_q^{(1)} = \omega_q^{(2)}$ ($\Delta=0$), while in panels (c)-(d), $2\Delta = \omega_q^{(1)} - \omega_q^{(2)} = 0.3 \omega_0$.
	\label{fig_trajectory1}}
\end{figure*}

Thus, let us analyze what occurs when a quantum jump takes place through the four possible  dissipation channels (for the sake of brevity, we indicate them with their rates $\kappa$, $\gamma_{(1,2)}$, and $\gamma_C$). 
If there is a cavity jump $\kappa$, the wave function is projected onto the state $\ket{0, g, g}$. At this point, the system does not evolve anymore. This behaviour, shown in \figpanel{fig_trajectory1}{a}, can occur both in the presence of local and collective qubit dissipation for both identical and non-identical qubits. 

On the other hand, an excitation can be detected from one qubit $\gamma_{(1,2)}$ or via collective dissipation $\gamma_C$. 
In the case of a local jump for qubit 1 via $\gamma_1$, the wave function $\ket{\psi (t)}$ in \eqref{initial_state} is projected onto 

\be \label{qubit_jump} 
\ket{\phi} = \frac{ \hat C_1^+ \ket{\psi (t)}}{\mleft[ \brakket{\psi (t)}{\hat C_1^- \hat C_1^+}{\psi (t)} \mright]^{1/2}} = - i \ket{0, g, e} \, ,
\ee

i.e., qubit 2 (qubit 1) is instantly excited (de-excited).
Similarly, if   qubit 2 jumps the system ends up in $\ket{0, e, g}$.  In the case of a collective qubit jump, $\ket{\psi (t)}$ is projected onto 

\be \label{collective_qubit_jump}
\begin{split}
\ket{\chi^+} &= \frac{ \mleft[ \hat C_1^+ + \hat C_2^+ \mright] \ket{\psi (t)}}{\mleft[  \brakket{\psi (t)}{\mleft[ \hat C_1^- + \hat C_2^- \mright] \mleft[ \hat C_1^+ + \hat C_2^+ \mright]}{\psi (t)} \mright]^{1/2}}\\
&=\frac{- i}{\sqrt{2}} \bigg ( \ket{0, g, e} + \ket{0, e, g} \bigg ) \, .
\end{split}
\ee

\renewcommand{\arraystretch}{1.4}

\begin{table*}
    \centering
    \begin{tabular}{|>{\centering\arraybackslash}m{0.05\textwidth}|>{\centering\arraybackslash}m{0.05\textwidth}|>{\centering \arraybackslash}m{0.2\textwidth}|>{\centering \arraybackslash}m{0.2\textwidth}|>{\centering \arraybackslash}m{0.2\textwidth}|>{\centering\arraybackslash}m{0.2\textwidth}|}
    \hline
     & & ${\rm{\Delta}}=0$, $\delta \gamma=0$  & ${\rm{\Delta}}\neq 0$, $\delta \gamma=0$ &  ${\rm{\Delta}}= 0$, $\delta \gamma\neq0$ &  ${\rm{\Delta}}\neq  0$, $\delta \gamma\neq0$ \\ \hline
    \rotatebox[origin=c]{90}{$\gamma_C=0$}& {\centering$ \ket{\phi} $ }& JC-like oscillations by ${\rm{\Omega}}_{\rm eff}^{(2)}$ [\figpanel{fig_trajectory1}{b}]& No evolution $\hat{U}(t) \propto \hat{\mathbbm{1}}$ [\figpanel{fig_trajectory1}{c,d}]& Non-sinusoidal oscillations if $\delta \gamma< {\rm{\Omega}}_{\rm eff}^{(2)}$, exponential decay otherwise (not shown)& $\hat{U}(t) \not\propto \hat{\mathbbm{1}}$, $\ket{\phi}$ is eigenstate of $\hat{U}(t)$: no evolution (not shown) \\ \hline
    \parbox[b]{0.05\textwidth}{\multirow{2}{*}{\rotatebox[origin=c]{90}{$\gamma_C\neq0$ \qquad}}}& {\centering$ \ket{\phi} $ }& Damped JC-like oscillations around the Bell state $(\ket{0, g, e} - \ket{0, e, g} )/\sqrt{2}$  [\figpanel{fig_trajectory3}{a}]& Competition between ${\rm{\Delta}}$ and $\gamma_C$ generates damping or small oscillations (not shown) & Competition among $\gamma_C$, $\delta \gamma$, and ${\rm{\Omega}}_{\rm eff}^{(2)}$ generates damped JC-like oscillations towards a state different from the Bell state $(\ket{0, g, e} - \ket{0, e, g} )/\sqrt{2}$ (not shown)
    &  If $\gamma_1>\gamma_2$: negligible state transfer [\figpanel{App:fig_trajectory4}{b}].
    If $\gamma_1<\gamma_2$: dissipative state transfer induced by competition between ${\rm{\Delta}}$, $\delta\gamma$ and $\gamma_C$ (not shown here, see Ref.~\cite{MingantiPRA2021})\\ 
    \cline{2-6}
    & $\ket{\chi^+}$ & $\ket{\chi^+}$ is an eignestate of $\hat{U}(t)$: no evolution [\figpanel{fig_trajectory3}{b}]& Oscillation between the Bell states $(\ket{0, g, e} \pm \ket{0, e, g} )/\sqrt{2}$ (not shown) & $\ket{\chi^+}$ is not an eigenstate of $\hat{U}(t)$: continuous undamped oscillations around the Bell state $(\ket{0, g, e} - \ket{0, e, g} )/\sqrt{2}$ (not shown) & $\delta \gamma$ favors either $\ket{0,g,e}$ or $\ket{0,e,g}$, the dynamics depending on $\delta \gamma$, $\gamma_C$, and ${\rm{\Delta}}$ [examples are given in \figpanel{fig_trajectory3}{c,d}]\\
    \hline
    \end{tabular}
    \caption{Evolution of the system in the  qubit-subspace  $\{\ket{0,g,e}$,$\ket{0,e,g}\}$ as it stems from \eqref{evolution_operator2}.
    $ \ket{\phi} $ represents the initial state for the evolution in the qubit-qubit excitation manifold after a local quantum jump of $\gamma_1$, see \eqref{qubit_jump}. $ \ket{\chi^+} $ is the initial state for the evolution in the qubit-qubit excitation manifold when a collective jump $\gamma_C$ occurs, see \eqref{collective_qubit_jump}.
    }
    \label{tab:my_label}
\end{table*}

\subsubsection{Qubit-qubit interaction}\label{sec:Qubit-qubit interaction}

By just collecting the first quantum jump of the system, one cannot know if the expected simultaneous qubit excitation takes place.
Indeed, it is not only necessary to detect (at a given time $t$) one excitation coming out from  qubit 1, but to be sure that qubit 2 was also excited at the same time as the quantum jump occurred. 

As such, the analysis of the dynamics between the first and the second quantum jumps allows to reconstruct all those sub-processes which take place when there is an excitation emitted from the qubits.
Note that the only possible states after the first quantum jump are $\ket{0,g,g}$, $\ket{0,e,g}$, $\ket{0,g,e}$ or $ \mleft ( \ket{0, g, e} + \ket{0, e, g} \mright)/\sqrt{2}$. So this demonstrates that the qubits have been excited simultaneously.
Thus, one needs to correctly describe the dynamics after the first quantum jump to characterize the second quantum jump taking place.
In the three cases where the dissipation occurs via the qubits ($\gamma_{(1,2)}$ or $\gamma_C$), each jump is followed by a new dynamics.
Notably, this occurs into the two-dimensional subspace $\{ \ket{0, e, g} , \ket{0, g, e} \}$, because the loss of one quibit excitation makes it impossible to excite back the cavity. 
The Hamiltonian part of the evolution is captured by $\hat{H}^{(2)}_{\rm eff}$, and we recall that ${\rm{\Omega}}^{(2)}_{\rm eff}=0$ for ${\rm{\Delta}}\neq0$ (see \appref{EffHamiltonian}). As such, we obtain:
\begin{widetext}
\be 
\begin{split}
\label{evolution_operator2} 
\hat U(t) &= e^{- \frac{1}{4} {\rm{\Gamma}} t}  \Bigg\{ \mleft[ \cos (\zeta t/4) - \frac{\delta \gamma + i {\rm{\Delta}} }{\zeta} \sin (\zeta t / 4) \mright] \ketbra{0, e, g}{0, e, g}  \\
&  - i \frac{4 {\rm{\Omega}}_{\rm eff}^{(2)} - i \gamma_C}{\zeta} \sin (\zeta t / 4) \bigg[ \ketbra{0, e, g}{0, g, e} + \ketbra{0, g, e}{0, e, g} \bigg]  
+  \mleft[ \cos (\zeta t / 4) + \frac{\delta \gamma + i {\rm{\Delta}} }{\zeta} \sin (\zeta t / 4) \mright] \ketbra{0, g, e}{0, g, e} \Bigg\} \, ,
\end{split}
\ee
\end{widetext}
where $\zeta = \sqrt{(4 {\rm{\Omega}}_{\rm eff}^{(2)} - i \gamma_C)^2 - (\delta \gamma + i {\rm{\Delta}})^2}$ is the new parameter determining the oscillation frequency,
and $\delta \gamma = \gamma_1 - \gamma_2$. Similarly to the previous case, the non-Hermitian evolution operator $\hat U(t)$ captures the dynamics before a quantum jump takes place.
Note that, since we are in the manifold where only the qubits are excited, the photon dissipation plays  no role.
No matter which quantum jump occurs, this time the system ends in $\ket{0,g,g}$. 

Having detailed the most general possible dynamics after the first quantum jump, several different types of behavior can take place, as detailed in Table~\ref{tab:my_label}. Although all of them are interesting to analyze, demonstrating the different effects of the environment, hereafter we focus on particular cases to efficiently characterize the presence of the one-photon--two-atom process. This complete landscape of the possible behavior of the system allows to characterize the emission of the system, so that one can undoubtedly know if the simultaneous excitation of the two atoms by a photon has taken place in an experiment.

In the following sections, we will numericaly simulate the full Hamiltonian and dissipative dynamics to prove the validity of this analysis. A detailed analytical derivation of these results (using the effective Hamiltonian) is provided in the \appref{AnalyticalI}.

\section{Results I: Single trajectories without collective dissipation}
\label{ResultI}

We now investigate the signatures of USC in the emission spectrum by considering the simplest case, where only the local dissipation $\gamma_{(1,2)}$ can act ($\gamma_C = 0$).

\subsection{One photon  exciting two atoms}

In Fig.~\ref{fig_trajectory1} and in the discussion below, we always initialize the system in $\ket{1,g,g}$ and we consider the case where the sum of the energy of the two qubits is resonant with the energy of the single photon ($\omega_c \simeq \omega_q^{(1)}+ \omega_q^{(2)} $).
As such, an oscillation where one photon excites two qubits occurs (as seen from all the panels in Fig.~\ref{fig_trajectory1}).
This oscillation is well captured by \eqref{oscillate_state_cavity}, as described in \secref{One-photon--two-atoms}.
After the initial evolution takes place, sooner or later, a quantum jump occurs. If it is a photon emission, the wave function is projected onto $\ket{0, g, g}$, where $\expec{\hat X^- \hat X^+} = \expec{\hat C_i^- \hat C_i^+} = 0$ as shown in \figpanel{fig_trajectory1}{a}.
If, instead, $\gamma_1$ or $\gamma_2$ occurs, the wave function $\ket{\psi (t)}$ [see \eqref{initial_state}] is projected onto the new initial normalized state $\ket{\phi} = - i \ket{0, g, e}$ or onto $\ket{\phi} = - i \ket{0, e, g}$. 
For the qubit emission, the new dynamics taking place between the two qubits is due to the second-order processes, as described in \secref{sec:Qubit-qubit interaction}.

\subsubsection{Histogram of quantum jumps}

\begin{figure*}
    \centering
    \includegraphics[width=.9\linewidth]{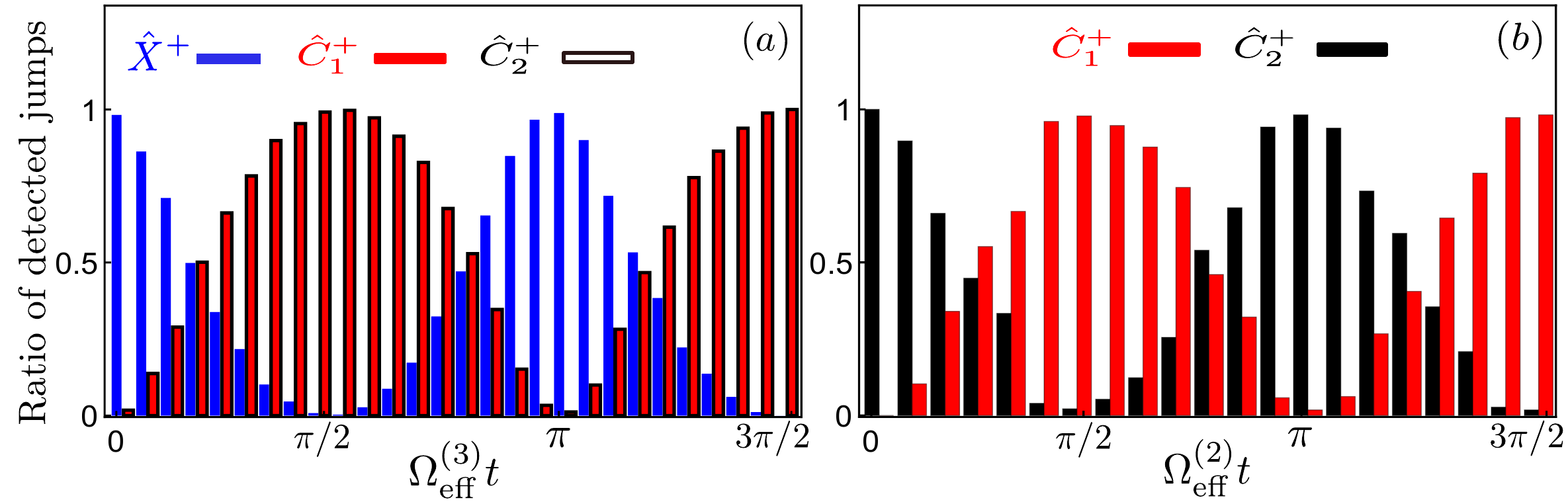}
    \caption{Histograms of the ratio of the total local quantum jumps as a function of time.
    (a) Local quantum jumps due to $\hat{X}^+$ (the cavity, blue bars), $\hat{C}_1^{+}$ (qubit 1, red bars), and $\hat{C}_2^{+}$ (qubit 2, unfilled black bars surrounding the red ones). The histogram is constructed from simulations of $2 \times 10^5$ trajectories. The system was initialized in the state $\ket{1, g, g}$.
    (b) Local quantum jumps due to $\hat{C}_1^{+}$ (qubit 1, red bars) and $\hat{C}_2^{+}$ (qubit 2, black bars) after an initial qubit 1 jump. The histogram was constructed from simulations of $4 \times 10^5$ trajectories and reveals oscillations like those in \figpanel{fig_trajectory1}{b}.
    Parameters for both panels: $\omega_q^{(1,2)} = \omega_0$ ($\Delta=0$), $\omega_c \simeq 2 \omega_0$, $g = 0.1 \omega_0$, $\kappa = \gamma_{(1,2)} = 4 \times 10^{-5} \omega_0$, and $\gamma_C = 0$.
    \label{fig:local_dissipation}}
\end{figure*}

Let us detail how the statistics of quantum jumps allows to witness the behaviours described in the preceding section. Although recording all the quantum jumps and then post-selecting trajectories corresponding to certain processes is possible, e.g., in experiments with superconducting circuits~\cite{NaghilooNatPhys19}, it can be difficult, not only because the energy transfer is a rare event, but especially when dealing with collective jumps. As discussed in Ref.~\cite{MingantiPRA2021}, a simple way to enable observation of all the processes of interest is to reconstruct them by creating histograms showing the distribution of the \emph{local} quantum jumps as a function of time.

To observe the one-photon--two-atom excitation process, we need to collect all the local quantum jumps from the cavity ($\mathcal{D}[\hat{X}^{+}]$) and from either of the two qubits ($\mathcal{D}[\hat{C}^+_1]$ and $\mathcal{D}[\hat{C}^+_2]$). Such a reconstruction of the process is shown in \figpanel{fig:local_dissipation}{a}. The characteristics of the energy exchange can be determined up to arbitrary precision by collecting enough data. Note that for finite-efficiency detectors that fail to detect some jumps, the overall jump statistics is unaffected, since on average the same amount of quantum jumps will be missed from the cavity and from the qubits. Reaching the wanted precision thus simply requires a higher number of realizations for worse detectors.

Before considering the second-order processes, let us motivate on a mathematical ground why such a procedure of collecting the quantum jumps allows to describe the dynamics.
Since the initial state $\ket{1, g,g}$ is $\hat{X}^{-} \ket{0,g,g}$, by considering the dynamics of the first jump we are witnessing the two-time correlation functions of the effective Hamiltonian of the system.
For example, the emission of a cavity quantum jump at time $t$ can be described as
\begin{equation}
   \bra{0,g,g} \hat{X}^{+}(t)\hat{X}^{-}(0) \ket{0,g,g} 
\end{equation}
This is the definition of the two-time correlation function of the real photon detection when no quantum jumps occur [the blue bars in \figpanel{fig:local_dissipation}{a}].
Although this is not exactly the Hamiltonian of the non-dissipative process, by appropriately determining the dissipation rates one can simulate the closed system, as we also argued using \eqref{oscillate_state_cavity}.

\subsection{Identical qubits}
\label{local_identical}

The system eventually undergoes a quantum jump. It either emits a photon, ending in $\ket{0,g,g}$ as shown in \figpanel{fig_trajectory1}{a}, or the quantum jump leads the system to $\ket{0, g, e}$ or $\ket{0, e, g}$  in \figpanel{fig_trajectory1}{b-d}.
What now strongly depends on the parameters is how the evolution takes place in the qubit-qubit excitation manifold (see \secref{sec:Qubit-qubit interaction} and Table~\ref{tab:my_label}).

If we consider  identical qubits, $\hat H_{\rm eff}^{(2)}$ in \eqref{Heff_second_oreder} is nonzero and
$\ket{0, g, e}$ (or equivalently $\ket{0, e, g}$) is not an eigenstate of the Hamiltonian. 
The cavity cannot be repopulated because there is not enough energy in the system. 
Nevertheless, the remaining energy continues to be exchanged between the qubits until a qubit jump occurs and the system wave function is projected onto the state $\ket{0, g, g} $ (see \appref{AnalyticalI}). For example, $\gamma_1$ emits a second time in \figpanel{fig_trajectory1}{b}.

\subsubsection{Histograms of the qubit quantum jumps}

To observe the excitation exchange between two resonant qubits that follows when a local qubit jump occurs, we need to be careful about how the quantum jumps are detected. If we were to simply create a histogram of the time distribution of the second quantum jump, we would not capture this phenomenon, since this procedure would reproduce the LME, which does not show the oscillations between the qubits (see \appref{App:comparison_dynamics}). Instead, the procedure to obtain the correct histogram is:

\begin{enumerate}
\item Monitor the dynamics until the first quantum jump takes place.
\item If the monitored event is a local quantum jump from one of the two qubits, restart the clock.
\item Monitor from which qubit the second jump takes place.
\item Collect data and make the histogram for the second quantum jump.
\end{enumerate}
Suppose that at a time $t$ there is a jump of $\hat{C}^{+}_{2}$. As such, the wave function collapses onto $\ket{0, e,g}$. Mathematically, we have
\begin{equation}
    \ket{0, e,g}=\hat{C}^{-}_{1} \ket{0,g,g}
\end{equation}
In other words, the procedure of monitoring the time $t+\tau$ when the second quantum jump occurs is equivalent to
\begin{equation}
   \bra{0,g,g} \hat{C}^{+}_{j}(t+\tau)\hat{C}^{-}_1(0) \ket{0,g,g} 
\end{equation}
This again is a well-defined two-time correlation function describing a non-Hermitian Hamiltonian evolution, whose characteristics can be obtained from \eqref{evolution_operator2}.

We plot the results of this histogram procedure in Fig.~\figpanelNoPrefix{fig:local_dissipation}{b}. We focus on those events where the first jump is caused by $\hat{C}_1^{+}$. We see that a periodic exchange of an excitation between the qubits takes place at a rate given by ${\rm{\Omega}}_{\rm eff}^{(2)}$. This process is much faster than the oscillation between the cavity and the two qubits (for our parameters, ${\rm{\Omega}}_{\rm eff}^{(2)} \simeq 10 {\rm{\Omega}}_{\rm eff}^{(3)}$), and thus requires the time bins to be much shorter than in \figpanel{fig:local_dissipation}{a}. Furthermore, the qubit-qubit oscillations only occur in a subset of all processes. Thus, with respect to the case shown in \figpanel{fig:local_dissipation}{a}, one is required to repeat the experiment more times in order to obtain sufficient statistics to generate  \figpanel{fig:local_dissipation}{b}. From it, we obviously can reconstruct the oscillations in \figpanel{fig_trajectory1}{b}.

The two histograms in Fig.~\ref{fig:local_dissipation} allow to reconstruct both the amplitude and the frequency of the oscillations. Combined toghether, not only do they demonstrate that a single photon excite the two atoms, but they also show the dynamics between the two qubits as part of the main effect.
This dynamics (enabled by a quantum jump) is completely missed by other protocols (see \appref{sec:Homodyne}) or by the averaging process of the Lindblad master equation (hidden by the averaging processes as in \appref{App:comparison_dynamics}). Signatures of such an oscillation could not be witnessed starting from $\ket{1,g,g}$, but would require to initialize the system in $\ket{0,e,g}$ or $\ket{0,g,e}$ (see \appref{comparison}).

\subsection{Non-identical qubits}
\label{local_nonidentical}
For non-identical qubits (${\rm{\Delta}} > {\rm{\Omega}}_{\rm eff}^{(2)}$), the second-order effective terms $\hat H_{\rm eff}^{(2)}$ in \eqref{Heff_second_oreder} can be neglected thanks to the RWA. 
Although a nontrivial dynamics can occur in this manifold due to the different decay rates of $\ket{0,g,e}$ and $\ket{0,e,g}$ [cf. \eqref{evolution_operator2}], this effect cannot be witnessed along a single quantum trajectory, since after a quantum jump the state will never be a superposition of $\ket{0,g,e}$ and $\ket{0,e,g}$ due to the nature of the local quantum jumps (see the discussion in \secref{single_trajectory_collective_dissipation}, where a jump of $\gamma_C$ will, instead, unveil this effect).
Thus, the qubits cannot exchange the remaining excitation anymore. 
This process is shown in \figpanel{fig_trajectory1}{c,d}, where a quantum jump first takes place in qubit 2 (qubit 1) and then in the other qubit. 

\subsubsection{Histogram of the quantum jumps}

As we previously stated, in the case of different qubits no exchange of excitations takes place between the two qubits. This fact can be used as a immediate witness of the simultaneous excitation of the two qubits by a single photon.
Indeed, once one of the qubit emits, for instance quibt 1, the state of the system remains in $\ket{0,g,e}$.
Therefore, the only possible event is an emission from qubit 2. In this case, we expect that the number of qubit jumps is identical for qubit 1 and qubit 2 independently of the dissipation rates $\gamma_1$ and $\gamma_2$. This indirectly demonstrates that the photon is simultaneously exciting both atoms. 
A more detailed time analysis reveals that the first quantum jump occurs more frequently in the more dissipative qubit, and then the less dissipative qubit follows.
We numerically verified this analysis (not shown here), and we stress that this prediction is true only for $\gamma_C=0$.

\section{Results II: Single trajectories considering both local and collective dissipation}\label{single_trajectory_collective_dissipation}

\begin{figure*}
	\centering
	\includegraphics[width=0.9\linewidth]{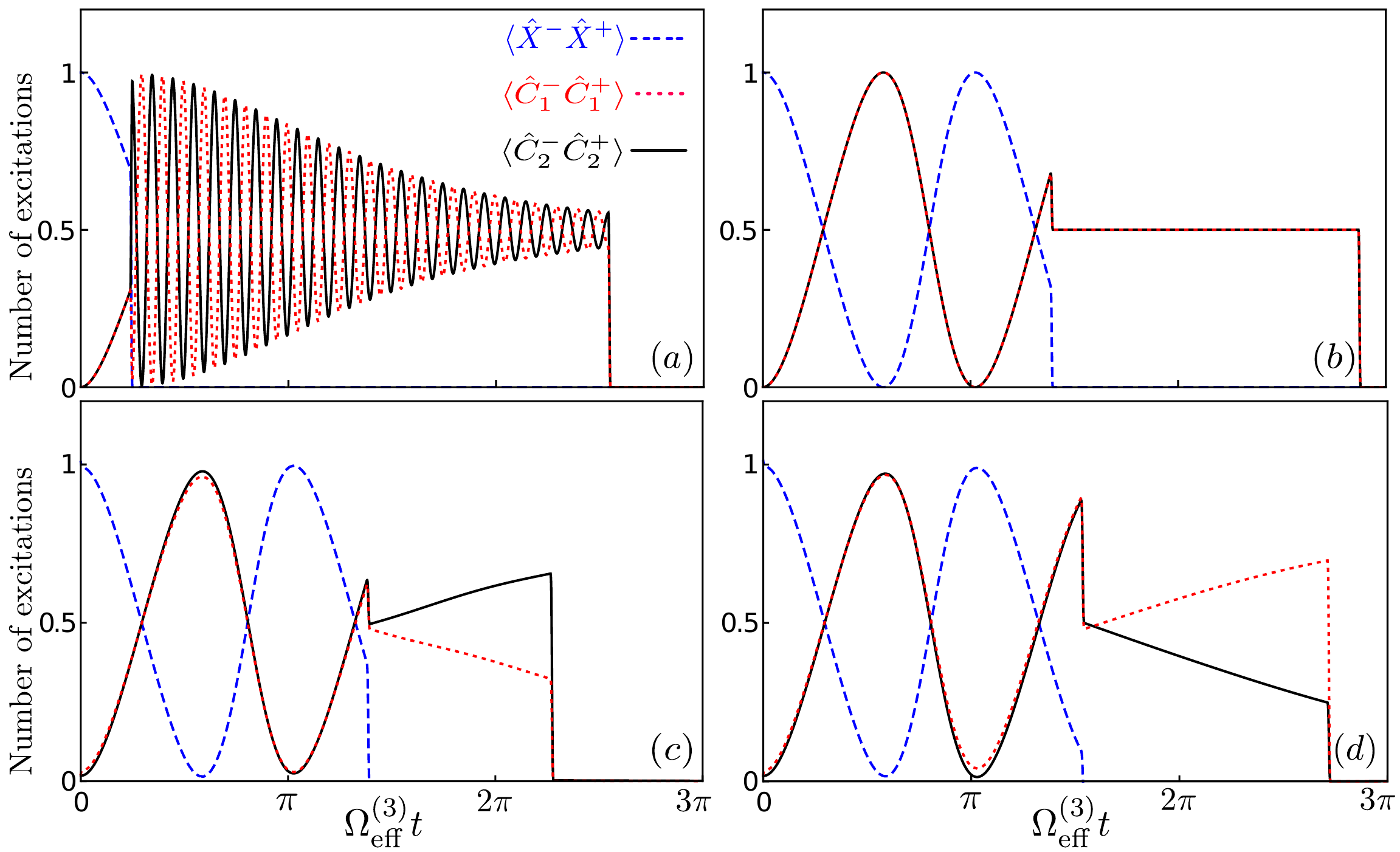}
	\caption{Examples of single quantum trajectories, numerically analyzed using the full Hamiltonian in \eqref{Hamiltonian} and the dissipators in  \eqref{Jump_operators} in the presence of collective qubit decay ($\gamma_C\neq 0$).
	All panels show the time evolution of the mean photon number $\expec{\hat X^- \hat X^+}$ (blue dashed curves) and of the mean excitation numbers of the two qubits $\expec{\hat C_i^- \hat C_i^+}$ ($i = 1, 2$) (red dotted and black solid curves, respectively).
    The system is always initialized in $\ket{1,g,g}$ and all the panels, starting in the resonant condition of the one-photon--two-atom process, initially display the oscillation in \eqref{initial_state} until a quantum jump occurs. The panels represent a quantum trajectory where
	(a) for identical qubits case, a qubit 1 jump occurs projecting the wave function onto the excited state of qubit 2. The two qubits start to exchange their excitation around the superposition state $\ket{\chi^{-}} =\mleft( \ket{0, g, e} - \ket{0, e, g} \mright)/\sqrt{2}$, slowly converging towards this state until another qubit jump occurs (local or collective), projecting the system onto the state $ \ket{0, g, g}$ (as discussed in Sec.~\ref{Collective_identical});
	(b) For identical qubits case, a collective qubit jump occurs projecting the wave function onto the superposition state $ \ket{\chi^+}= -i \mleft( \ket{0, g, e} + \ket{0, e, g} \mright)/\sqrt{2}$. Despite this being the ``bright'' state (c.f. Sec.~\ref{Collective_identical}), the system remains in this state until another jump occurs, projecting the system onto the state $ \ket{0, g, g}$;
	(c) For non-identical qubits case, a collective qubit jump occurs with non-identical relaxation rates as discussed in Sec.~\ref{Collective_nonidentical}. Since $\gamma_1 = 4 \times 10^{-4} \omega_0 > \gamma_2 = 4 \times 10^{-5} \omega_0$, the probability of measuring qubit 1 (qubit 2) in its excited state decreases (increases) as time increases until another jump occurs, projecting the system onto the state $ \ket{0, g, g}$.
	(d) For non-identical qubits case, a collective qubit jump occurs  with non-identical relaxation rates but with the values of $\gamma_1$ and $\gamma_2$ interchanged with respect to (c), leading to the opposite process.
	In all panels, the parameters  are $g = 0.1 \omega_0$, $\omega_c \simeq 2 \omega_0$, $\kappa = 4 \times 10^{-5} \omega_0$, and $\gamma_C = 5 \times 10^{-4} \omega_0$. In panels  (a)-(b) $\omega_q^{(1,2)}=\omega_0$ ($\Delta=0$) while in (c)-(d), ${2\rm{\Delta}}=\omega_q^{(1)}-\omega_q^{(2)} = 0.3 \omega_0$.
	\label{fig_trajectory3}}
\end{figure*}

The previous section undoubtedly demonstrate the presence of the main one-photon--two-atoms process in the case $\gamma_C=0$. However, in actual experimental realization, collective dissipation naturally emerges due to the coupling of the qubits with a common environment. 
In this case, although the evolution operator for the one-photon--two-atoms in \eqref{evolution_operator} depends on $\gamma_C$, the plots in Fig.~\ref{fig_trajectory3} do not significantly deviate from those in Fig.~\ref{fig_trajectory1} for small-enough $\gamma_C$.
As we detail below, such a coupling deeply changes the characteristics of the qubit-qubit dynamics, requiring a different analysis if the two qubits can lose their excitations via the collective jump operator $\sqrt{\frac{\gamma_C}{2}} (\hat C_1^+ + \hat C_2^+)$. 
We again study the cases of identical and non-identical qubits separately.

\subsection{Identical qubits}\label{Collective_identical}

After a local qubit jump, as in \figpanel{fig_trajectory3}{a}, the cavity cannot be repopulated and the two qubits start exchanging an excitation as in the case of \figpanel{fig_trajectory1}{b}.
However, differently from the other case, the collective dissipation forces the system toward the superposition state $\ket{\chi^{-}}=(\ket{0, g, e} - \ket{0, e, g})/\sqrt{2} $ until a collective or local qubit jump occurs, projecting the wave function onto the state $\ket{0, g, g}$. Such a peculiar behavior can be argued from the action of the three dissipation channels.
Indeed, any state in the qubit-qubit manifold can be described via the superposition of Bell states  $\ket{\chi^{\pm}}=(\ket{0, g, e} \pm \ket{0, e, g})/\sqrt{2} $.
While the local dissipations $\gamma_{(1,2)}$ act identically on $\ket{\chi^{\pm}}$, the collective one does not affect the evolution of $\ket{\chi^{-}}$. As such, the presence of $\gamma_C$ forces the system onto the ``dark state'' $\ket{\chi^-}$, because the ``bright'' state $\ket{\chi^{+}}$ decays more rapidly even when quantum jumps do not occur (see the discussion in \appref{AnalyticalI} and in Ref.~\cite{MingantiPRA2021}).
Thus, no matter the details of the initial state, the wave function tends towards the superposition state $\ket{\chi^{-}}$.

When the collective dissipation acts, the wave function is instead projected onto the superposition state proportional to $\ket{\chi^{-}}$ [see \eqref{collective_qubit_jump}].
Since $\ket{\chi^{-}}$ is an eigenstate of the effective Hamiltonian $\hat H_{\rm eff}$ in \eqref{Heff0},
the state does not evolve with \eqref{evolution_operator2}, as shown in \figpanel{fig_trajectory3}{b}.

\subsubsection{Histograms in the presence of collective dissipation}

\begin{figure*}
    \centering
    \includegraphics[width=0.75
    \linewidth]{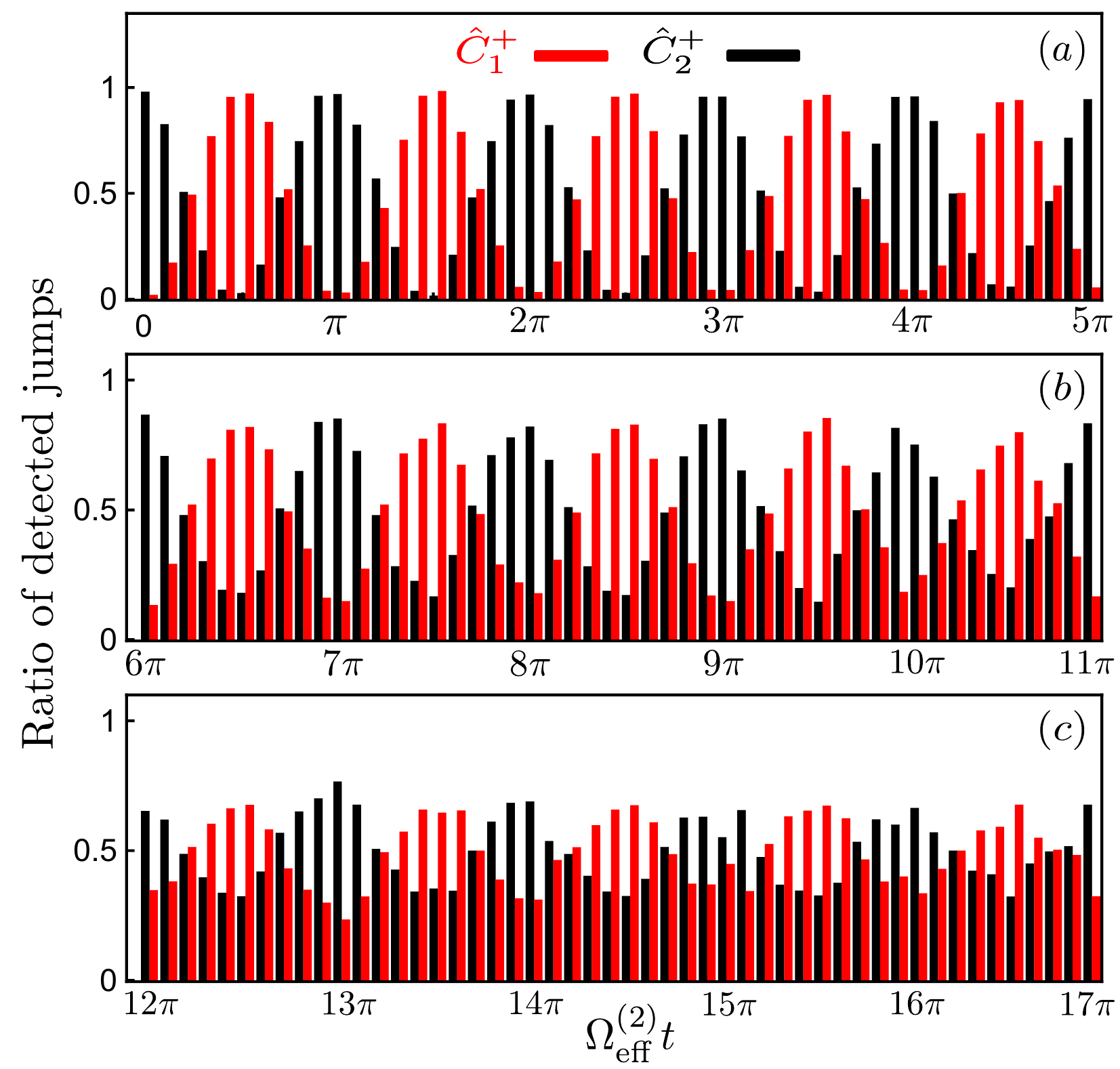}
    \caption{Histograms of the ratio of the total local quantum jumps as a function of time, due to $\hat{C}_1^{+}$ (qubit-1, red bars) and $\hat{C}_2^{+}$ (qubit 2, black bars), after an initial qubit 1 jump. As time progresses from panel (a) to panel (c), the oscillation amplitude decreases reaching the superposition state $\ket{\chi^{-}} =\mleft( \ket{0, g, e} - \ket{0, e, g} \mright)/\sqrt{2}$ . The histograms are constructed from simulations of $8 \times 10^5$ trajectories and reconstruct dynamics similar to \figpanel{fig_trajectory3}{a}.
    Parameters are the same as in \figref{fig:local_dissipation}, except for $\gamma_C = 5 \times 10^{-4}\omega_0$.
    \label{fig:no_local_dissipation}}
\end{figure*}

Even in the presence of collective dissipation, we can use the histogram of the local quantum jumps to characterize the phenomena taking place.
While the one-photon--two-atoms process remains almost identical, the qubit excitation exchange is affected by $\gamma_C$ as just described.
Using the same procedure as in \secref{local_identical}, we can again obtain the two-time correlation functions allowing to witness the presence of the damped-oscillation behavior.
This is plotted in \figref{fig:no_local_dissipation}, where we see that the qubit-qubit oscillations gradually decrease in amplitude towards the value $1/2$. This is due to the system converging to the Bell state $\ket{\chi^{-}}$ (see the discussion in the \appref{AnalyticalI}).

Although for collective dissipation we only plot cases in which the two qubits were exactly on resonance, the technique demonstrated here can also be applied to cases where ${\rm{\Delta}} \neq 0$ and $\gamma_1 \neq \gamma_2$. 
In the latter case, the competition between these several processes can induce interesting behaviours, whose discussion goes beyond the purpose of this article, requiring a detailed study of the competing ratios.
We refer the interested reader to Ref.~\cite{MingantiPRA2021}, and we note that anyhow these histograms can be used to extract the effective couplings ${\rm{\Omega}}_{\rm eff}^{(2)}$ and ${\rm{\Omega}}_{\rm eff}^{(3)}$ as well as the collective dissipation rate $\gamma_C$.

\subsection{Non-identical qubits}
\label{Collective_nonidentical}
For non-identical qubits the second-order effective terms $\hat H_{\rm eff}^{(2)}$ in \eqref{Heff_second_oreder} can be neglected. Although there is no Hamiltonian interaction, the presence of the collective dissipation enables a non-Hermitian coupling between the qubits  (see the discussion in \appref{AnalyticalI} and Ref.~\cite{MingantiPRA2021}).
Mathematically, this can be seen by the action of the off-diagonal terms of the time evolution operator activated by $\gamma_C$, as it stems from \eqref{evolution_operator2}. 

At first, let us consider the collective-qubit jump case where the wave function after the jump $\ket{\chi^{-}}$ [see \eqref{collective_qubit_jump}] evolves as
\begin{widetext}
\be\label{second_superposition_state_collective_non_identical}
\begin{split}
\ket{\psi (t)} &= - \frac{i e^{- \frac{1}{4} {\rm{\Gamma}} t} }{\sqrt{2}} \mleft\{ \mleft[ \cos (\zeta t / 4) - \frac{\gamma_C}{\zeta} \sin (\zeta t / 4) \mright] \bigg [ \ket{0, e, g} + \ket{0, g, e} \bigg ] \mright. - \mleft. \frac{\delta \gamma + i {\rm{\Delta}} }{\zeta} \sin (\zeta t / 4) \bigg [ \ket{0, e, g} - \ket{0, g, e} \bigg ] \mright\} \, ,
\end{split}
\ee
\end{widetext}
where ${\rm{\Gamma}} = \gamma_1 + \gamma_2 + \gamma_C$  and $\zeta = \sqrt{({\rm{\Delta}}  - i \delta \gamma)^2 - \gamma_C^2}$.

For $\gamma_1 = \gamma_2$ (and sufficiently small $\gamma_C$), $\zeta$ is real and the state $\ket{\chi^+}$ oscillates between the two Bell states $\ket{\chi^{\pm}}$ (see Table~\ref{tab:my_label}). Indeed, the effect of ${\rm{\Delta}}$ can be seen as a term inducing a rotation of the Bell states. By selecting the correct $\delta \gamma$ one can fix the initial state that remains in the initial superposition state $\ket{\chi^{-}}$, until a jump projects the wave function onto the state $\ket{0,g,g}$.
Otherwise, the state oscillates but the expectation values remain constant and $\expec{\hat C_1^- \hat C_1^+}=\expec{\hat C_2^- \hat C_2^+}=1/2$ as shown in \figpanel{fig_trajectory3}{b}.

In the case $\gamma_1 = \gamma_C \gg \gamma_2$, we find that the probability of measuring qubit 1 in its excited state decreases as time increases, while the probability of qubit 2 being in its excited state increases. This behavior is due to the difference between the loss rates, which imply that, if no jump occurs, it is more likely for the system to be in the excited state of qubit 2, since that state has a lower probability of leading to a jump, as shown in \figpanel{fig_trajectory3}{c,d}. A more detailed analytical discussion can be found in \appref{AnalyticalI}. 

\section{Conclusion and outlook}
\label{sec:ConclusionOutlook}

We have shown how to apply the theory of quantum trajectories to systems with ultrastrong coupling between light and matter. This has at least two applications. Firstly, we can now obtain the time evolution of dissipative ultrastrongly coupled systems by averaging over the stochastic wave functions of several quantum trajectories instead of using a Lindblad master equation for the system density matrix. In some cases, the quantum-trajectory method is preferable to use, since the density matrix dynamics requires more computer resources than the wave function one.

The second application of quantum trajectories for ultrastrongly coupled systems is that individual trajectories can reveal behaviours of the system, connected to measurement back-action, that are hidden by the averaging inherent in a master-equation approach. We illustrated this for the setup in Ref.~\cite{Garziano2016}, where two atoms (qubits) are ultrastrongly coupled to a cavity mode. 

When the energy of the two qubits sum up to the energy of a single photon in the cavity, the USC enables a process where the system state oscillates back and forth between having one photon in the cavity and having both qubits excited. By studying quantum trajectories where the system output is measured with photodetectors, we showed that if a quantum jump in one of the qubits is detected, the system dynamics switch from the oscillation between one photon and the two qubits to oscillation between the two qubits. 

We further studied the example with the one-photon--two-atom excitation process for the qubits on and off resonance with each other, with and without collective qubit dissipation. We showed how these different cases can modify the behaviour that the system displays after detecting a quantum jump from one of the qubits.

We also put forward an experimental protocol for observing the above-mentioned effects using photodetection. In such an experiment, which we believe is feasible using circuit QED, the output photon flux emitted by a resonator can be measured in a photodetection experiment, while qubit emission can be detected by coupling it to an additional microwave antenna~\cite{Hofheinz2009}.

Looking to the future, we hope that the theoretical methods presented herein will find applications in experiments on systems with USC that take advantage of individual measurements to characterize processes that otherwise are hidden by averaging. The literature contains many examples beyond that of Ref.~\cite{Garziano2016}, which was analyzed here. Furthermore, having the theoretical description of quantum trajectories should enable the development of feedback schemes that could control ultrastrongly coupled systems in new ways.

\begin{acknowledgments}

AFK acknowledges support from the Japan Society for the Promotion of Science (BRIDGE Fellowship BR190501), the Swedish Research Council (grant number 2019-03696), and from the Knut and Alice Wallenberg Foundation through the Wallenberg Centre for Quantum Technology (WACQT).
S.S. acknowledges the Army Research Office (ARO) (Grant No. W911NF-19-1-0065).
F.N. is supported in part by: Nippon Telegraph and Telephone Corporation (NTT) Research, the Japan Science and Technology Agency (JST) [via the Quantum Leap Flagship Program (Q-LEAP), the Moonshot R$\&$D Grant Number JPMJMS2061, and the Centers of Research Excellence in Science and Technology (CREST) Grant No. JPMJCR1676], the Japan Society for the Promotion of Science (JSPS) [via the Grants-in-Aid for Scientific Research (KAKENHI) Grant No. JP20H00134 and the JSPS-RFBR Grant No. JPJSBP120194828], the Army Research Office (ARO) (Grant No. W911NF-18-1-0358), the Asian Office of Aerospace Research and Development (AOARD) (via Grant No. FA2386-20-1-4069), and the Foundational Questions Institute Fund (FQXi) via Grant No. FQXi-IAF19-06. 
\end{acknowledgments}

\appendix

\section{Derivation of the effective Hamiltonian}
\label{EffHamiltonian}

In order to derive the effective Hamiltonian in \eqref{Heff0}, we start from \eqref{Hamiltonian} in  \secref{sec:EffectiveHamiltonian} of the main text, transforming it to the interaction picture we obtain
\be
\label{App:Hamiltonian}
\begin{split}
\hat H_{\rm I}(t) &= g \cos \theta \;\hat a^\dag \sum_{i=1}^2 \mleft[ \hat \sigma_-^{(i)} e^{i \mleft( \omega_c - \omega_q^{(i)} \mright) t} + \hat \sigma_+^{(i)} e^{i \mleft( \omega_c + \omega_q^{(i)} \mright) t} \mright] \\
& \quad \; + g \sin \theta\; \hat a^\dag \sum_{i=1}^2 \hat \sigma_z^{(i)} e^{i 2 \omega_0 t} + \rm H.c. \, ,
\end{split}
\ee
{\footnotesize $$\, $$}
where H.c.~denotes Hermitian conjugate. 

Taking the resonant cavity frequency $\omega_c = \omega_q^{(1)} + \omega_q^{(2)} = 2 \omega_0$ and defining $2 {\rm{\Delta}} = \omega_q^{(1)} - \omega_q^{(2)}$, such that $\omega_q^{(1)} = \omega_0 + {\rm{\Delta}}$ and $\omega_q^{(2)} = \omega_0 - {\rm{\Delta}}$, we can define the five operators
\be \label{hoperators}
\begin{split}
\hat h_1 e^{i \omega_1 t} =& g \cos \theta \; \hat a^{\dag} \hat \sigma_{-}^{1} e^{i (\omega_0 - {\rm{\Delta}}) t} \\
\hat h_2 e^{i \omega_2 t} =& g \cos \theta \;\hat a^{\dag} \hat \sigma_{-}^{2} e^{i (\omega_0 + {\rm{\Delta}}) t}  \\
\hat h_3 e^{i \omega_3 t} =& g \cos \theta \;\hat a^{\dag} \hat \sigma_{+}^{1} e^{i (3 \omega_0 + {\rm{\Delta}}) t} \\
\hat h_4 e^{i \omega_4 t} =& g \cos \theta \;\hat a^{\dag} \hat \sigma_{+}^{2} e^{i (3 \omega_0 - {\rm{\Delta}}) t} \\
\hat h_5 e^{i \omega_5 t} =& g \sin \theta\; \hat a^{\dag} \sum_{i=1}^2 \hat \sigma_z^{(i)} e^{i 2 \omega_0 t} \, .
\end{split}
\ee

In terms of these operators, the system Hamiltonian in \eqref{App:Hamiltonian} can be written as

\be \label{APP:Heff0} 
\hat H_{\rm I}(t) = \sum_{m=1}^5 \mleft[ \hat h_m e^{i \omega_m t} + \hat h_m^\dag e^{- i \omega_m t} \mright] \, .
\ee

We now apply the generalized James' effective Hamiltonian method~\cite{Shao2017} which at the second order gives

\be \label{APP:second-oredr} 
\hat H_{\rm I}^{(2)} (t) = \sum_{j,k} \frac{1}{\omega_k} \mleft[ \hat h_j \hat h_k^\dag e^{i (\omega_j - \omega_k) t} - \hat h_j^\dag \hat h_k e^{- i (\omega_j - \omega_k) t} \mright] \, ,
\ee

while at the third order it gives
\begin{widetext}
\be \label{app:third-order}
\begin{split}
\hat H_{\rm I}^{(3)}(t) = \sum_{i,j,k}  &\bigg[ 
\frac{\hat h_i \hat h_j^\dag \hat h_k e^{i (\omega_i - \omega_j + \omega_k) t} + \hat h_i^\dag \hat h_j \hat h_k^\dag e^{i (- \omega_i + \omega_j - \omega_k) t}  +  \hat h_i \hat h_j \hat h_k^\dag e^{i (\omega_i + \omega_j - \omega_k) t} + \hat h_i^\dag \hat h_j^\dag \hat h_k e^{i (- \omega_i - \omega_j + \omega_k) t} }{\omega_k (\omega_j - \omega_k)} \\
&+  \frac{
\hat h_i^\dag \hat h_j \hat h_k e^{i (- \omega_i + \omega_j + \omega_k) t} + \hat h_i \hat h_j^\dag \hat h_k^\dag e^{i (\omega_i - \omega_j - \omega_k) t} 
}{\omega_k (\omega_j + \omega_k)}  \bigg] \, .
\end{split}
\ee
\end{widetext}
In the RWA, all frequency contributions which are significantly different from zero can be neglected.
Since the frequencies $\omega_m$ are all different, we only keep the terms in $\hat H_{\rm I}^{(2)} (t)$ [$\hat H_{\rm I}^{(3)} (t)$] where the sum of any two (three) frequencies is zero both for identical (${\rm{\Delta}} = 0$) and non-identical (${\rm{\Delta}} \neq 0$) qubits cases.  

\subsection{Identical qubits}
\label{app:identical_qubits} 
For the identical-qubit case, the second-order effective Hamiltonian [\eqref{APP:second-oredr}]  in the interaction picture reads

\be \label{App:second-order_identical_qubits} 
\begin{split}
\hat H_{\rm I}^{(2)} &= - \frac{2 g^2 \cos^2\theta }{3\omega_0}\mleft(\hat \sigma_z^{(1)} + \hat \sigma_z^{(2)} \mright ) \mleft( \hat a^\dag \hat a + \frac{1}{2} \mright)\\
&\quad - \frac{g^2 \sin^2 (\theta)}{2 \omega_0}\mleft( \hat \sigma_z^{(1)} + \hat \sigma_z^{(2)} \mright)^2 \\
&\quad -\frac{4 g^2 \cos^2\theta }{3 \omega_0} \mleft(  \hat \sigma_-^{(1)} \hat \sigma_+^{(2)}  +  \hat \sigma_+^{(1)} \hat \sigma_-^{(2)} \mright) \, .
\end{split}
\ee
We are always interested in the processes between the states $\ket{1,g,g}$ and $\ket{0,e,e}$ (one-photon--two-atoms manifold) and those between $\ket{0,e,g}$ and $\ket{0,g,e}$ (qubit-qubit excitation manifold).

\subsubsection{One-photon--two-atoms manifold}

In the one-photon--two-atoms manifold, the first two terms dress the two bare states, inducing a positive shift in the energy of noninteracting eigenstates, $\ket{1, g,g}$ and a negative one for $\ket{0, e,e}$.
As such, 
\begin{equation}
\hat{H}_{\rm shift}^{(2)} = \frac{4 g^2 \cos^{2}\theta}{3\omega_0}\bigg (\ket{1, g,g}\bra{1, g,g} - \ket{0, e,e}\bra{0,e,e}\bigg )
\end{equation}
This diagonal shift can be always eliminated by appropriately tuning $\omega_c$ and $\omega_q^{(i)}$ in \eqref{Hamiltonian}. As such,  we do not report these terms in the main text, but we always specify $\omega_c \simeq 2 \omega_0$. The third term of \eqref{App:second-order_identical_qubits}, instead, plays no role because it is always zero in the one-photon--two-atoms manifold.

The main term leading to the one-photon--two-atom excitation exchange is described by the third-order effective Hamiltonian which, starting from \eqref{app:third-order} we obtain the Eqs.~(\ref{Heff_third_oreder}) and (\ref{Omega_3_identical}) in the main test, which reads
\be \label{app:identical_third-order} 
H_{\rm eff}^{(3)}= - \frac{8 g^3 \cos^2\theta \sin \theta}{3 \omega_0^2}
\mleft( \hat a \hat \sigma_+^{(1)} \hat \sigma_+^{(2)} + \hat a^\dag \hat \sigma_-^{(1)} \hat \sigma_-^{(2)} \mright) \, .
\ee

\subsubsection{Qubit-qubit manifold}

In the qubit-qubit excitation manifold, the first two terms of \eqref{App:second-order_identical_qubits} are always zero, while the last term is an effective Jaynes--Cummings-like qubit-qubit interaction, which in the main text [see Eqs.~(\ref{Heff_second_oreder}) and (\ref{Omega_2_identical})] we call

\be \label{app:second-order_identical_interaction} 
\hat H_{\rm eff}^{(2)} = - \frac{4 g^2 \cos^2\theta}{3 \omega_0} (\hat \sigma_-^{(1)} \hat \sigma_+^{(2)} + \hat \sigma_+^{(1)} \hat \sigma_-^{(2)}) \, .
\ee
The third-order term in \eqref{App:second-order_identical_qubits} plays no  role in this manifold.

Equations~(\ref{app:identical_third-order})~and~(\ref{app:second-order_identical_interaction}) yield the effective Hamiltonian

\be \label{total_effective_Hamiltonian_non_identical}
\hat H_{\rm eff} = \hat H_{\rm shift}^{(2)} + \hat H_{\rm eff}^{(2)} + \hat H_{\rm eff}^{(3)} \, ,
\ee
which is equivalent to \eqref{Heff0} in the interaction picture (see \secref{sec:EffectiveHamiltonian} in the main text).

\subsection{Non-identical qubits}
\label{app:non_identical_qubits} 

For the non-identical qubits (${\rm{\Delta}} \gg 4 g^2 \cos^2\theta / 3 \omega_0$), 
in the Schr\"{o}dinger picture \eqref{APP:second-oredr} is written as

\begin{widetext}
\be \label{App:second-order_shift}
\begin{split}
\hat H_{\rm I}^{(2)} =\hat H_{\rm shift}^{(2)} &= \mleft[ 2 \omega_0 - \frac{2 g^2 \cos^2\theta (\omega_0 + {\rm{\Delta}})}{(\omega_0 - {\rm{\Delta}}) (3 \omega_0 + {\rm{\Delta}})} \hat \sigma_z^{(1)} -\frac{2 g^2 \cos^2\theta (\omega_0 - {\rm{\Delta}})}{(\omega_0 + {\rm{\Delta}}) (3 \omega_0 - {\rm{\Delta}})} \hat \sigma_z^{(2)} \mright] \hat a^\dag \hat a -\frac{g^2 \sin^2 \theta}{2 \omega_0} \mleft( \hat \sigma_z^{(1)} + \hat \sigma_z^{(2)} \mright)^2 \\
& \qquad +\mleft[ \omega_0 + {\rm{\Delta}} - \frac{g^2 \cos^2\theta (\omega_0 + {\rm{\Delta}})}{(\omega_0 - {\rm{\Delta}}) (3 \omega_0 + {\rm{\Delta}})} \mright] \hat \sigma_z^{(1)} + \mleft[ \omega_0 - {\rm{\Delta}} - \frac{g^2 \cos^2\theta (\omega_0 - {\rm{\Delta}})}{(\omega_0 + {\rm{\Delta}}) (3 \omega_0 - {\rm{\Delta}})} \mright] \hat \sigma_z^{(2)}  \, .
\end{split}
\ee
\end{widetext}
Notice that the second-order effective Hamiltonian does not induce any coherent  resonant  coupling between  the  two  qubits, since they are out of resonance, but it still induces an energy shift (which can be compensated by an appropriate choice of the parameters).

Despite the absence of a coherent interaction between the qubits in the qubit-qubit interaction manifold, the main one-photon--two-qubit process can still take place. Indeed, from \eqref{app:third-order} we obtain

\be \label{app:non_identical_third-order} 
\begin{split}
\hat H_{\rm I}^{(3)} = H_{\rm eff}^{(3)} =& -\frac{8 g^3 \cos^2\theta \sin \theta (3 \omega_0^2 + {\rm{\Delta}}^2)}{(\omega_0^2 - {\rm{\Delta}}^2) (9 \omega_0^2 - {\rm{\Delta}}^2)} \\
&\qquad \mleft( \hat a \hat \sigma_+^{(1)} \hat \sigma_+^{(2)} + \hat a^\dag \hat \sigma_-^{(1)} \hat \sigma_-^{(2)} \mright)  \, .
\end{split}
\ee
Notice that \eqref{app:non_identical_third-order} recovers \eqref{app:identical_third-order} for ${\rm{\Delta}}=0$.

\section{Comparison of energy levels obtained using the effective and the full system Hamiltonian}
\label{comparison}

\begin{figure}
	\centering
	\includegraphics[width=1\columnwidth]{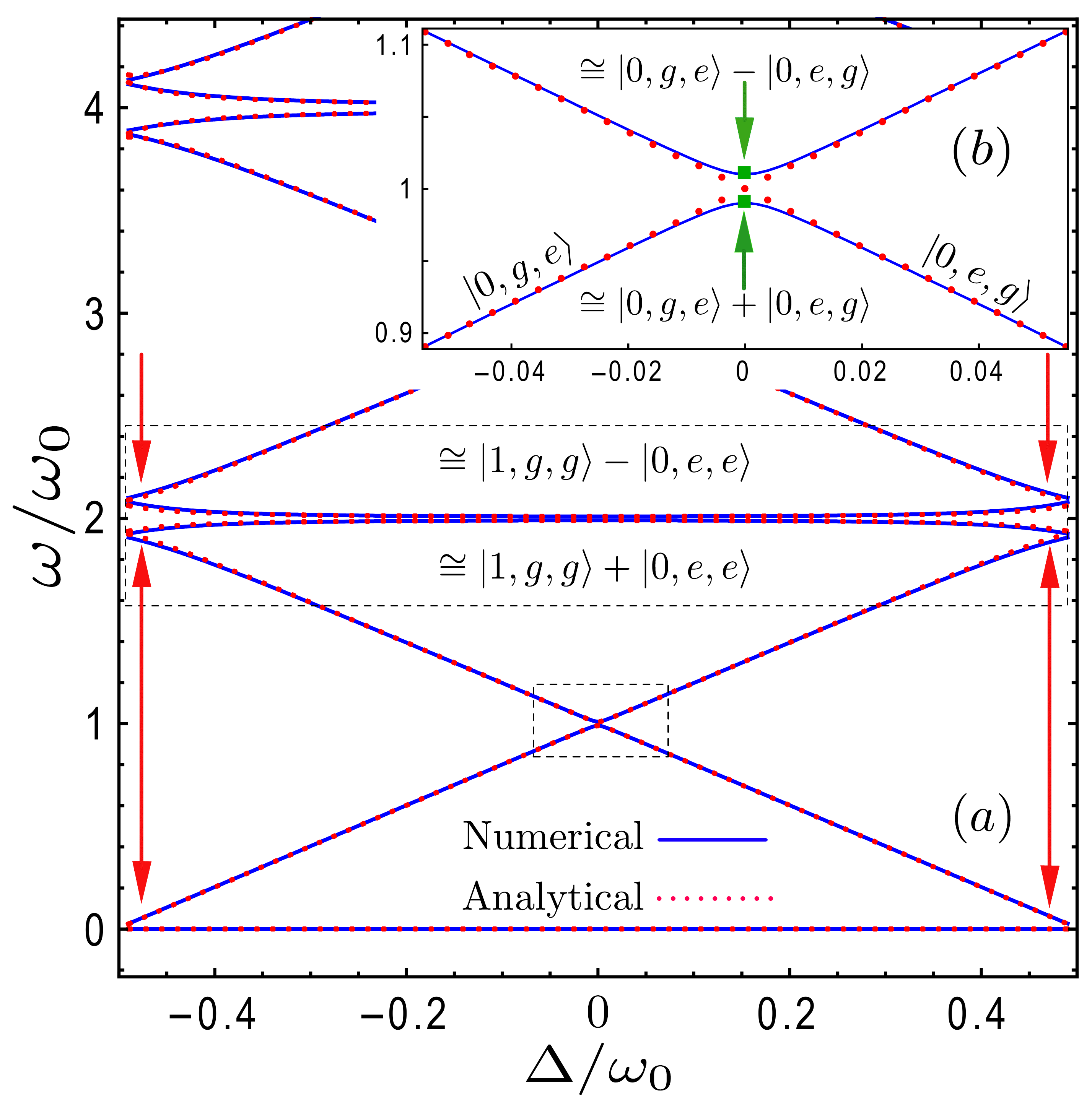}
	\caption{Energy levels for the system at the resonance $\omega_c \simeq \omega_q^{(1)} + \omega_q^{(2)}$ that enables the one-photon--two-atom excitation process.
	(a) Lowest energy levels as a function of ${\rm{\Delta}} / \omega_0$ of the full Hamiltonian from \eqref{Hamiltonian} (blue solid curves) and the effective Hamiltonians for identical (green dot) and non-identical (red dotted curves), obtained for $g / \omega_0 = 0.1$. The large dotted black rectangle delimits the region where the avoided level crossing (related to the one-photon--two-atom excitation process) appears. The red arrows indicate the limit of validity of the approximation. For large ${\rm{\Delta}}$, the coherent resonant coupling ${\rm{\Omega}}_{\rm eff}^{(3)}$ tends to become too small, so that the one-photon--two-atom excitation process becomes less likely.
	(b) An enlarged view of the first avoided level crossing [the small black dashed rectangle in panel (a)]. The avoided level crossing is due to the second-order effective interaction ${\rm{\Omega}}_{\rm eff}^{(2)}$ in \eqref{Heff_second_oreder} which is non-negligible only for ${\rm{\Delta}}=0$ (green squares).
	\label{App:fig_spectrum}}
\end{figure}
Here, we compare the lowest energy levels obtained for the effective Hamiltonian in \eqref{Heff0} with those calculated using the full system Hamiltonian in \eqref{Hamiltonian}. Figure~\figpanelNoPrefix{App:fig_spectrum}{a} shows the lowest energy levels of the full system Hamiltonian (blue solid curve) and those obtained by diagonalizing the effective Hamiltonian (red dotted curves), as a function of the frequency difference of the bare qubits ${\rm{\Delta}}$. The results are plotted for parameters fulfilling the resonance condition $\omega_c \simeq \omega_q^{(1)} + \omega_q^{(2)}$ and show excellent agreement. In the inset [\figpanel{App:fig_spectrum}{b}] an enlarged view of the first avoided level crossing [marked by a small black dashed rectangle in \figpanel{App:fig_spectrum}{a}] is shown. As expected, at its minimum (${\rm{\Delta}}=0$, green square) the energy difference between the Hamiltonian eigenstates is twice the effective resonant coupling ${\rm{\Omega}}_{\rm eff}^{(2)}$. This coupling is only important when the qubits are almost identical (${\rm{\Delta}} \approx 0$). For ${\rm{\Delta}} \gg {\rm{\Omega}}_{\rm eff}^{(2)}$, instead, the Jaynes--Cummings-like effective interaction is negligible due to the RWA (red dots), meaning that there is no longer a Hamiltonian coupling between the two qubits.  
The large dotted black rectangle in the centre of \figpanel{App:fig_spectrum}{a} delimits the region for which the one-photon-two-atom excitation process occurs. This region is quite large, meaning that, for various values of ${\rm{\Delta}}$ the coherent resonant coupling ${\rm{\Omega}}_{\rm eff}^{(3)}$ between the two states $\ket{1, g, g}$ and $\ket{0, e, e}$ does not change significantly. However, when the energies of the qubits becomes too different the coherent resonant coupling ${\rm{\Omega}}_{\rm eff}^{(3)}$ tends to be too small and the one-photon--two-atom excitation process becomes less likely. For even larger values, the energy of one quibt become comparable to that of the cavity, and the James' approximation breaks, as marked by the deviation of the red dots indicated by the red arrows.
\section{Analytical results}
\label{AnalyticalI}

Here, we carry out analytical calculations using the non-Hermitian Hamiltonian in \eqref{non_Hermitian_H} with the effective Hamiltonian in \eqref{Heff0}.  The main one-photon--two-atom process has already been described in the main text. Here, we focus on the second-order processes occuring in the qubit-qubit manifold once the first quantum jump took place. 
We recall that 
\begin{equation}
\begin{split}
    \zeta &= \sqrt{(4 {\rm{\Omega}}_{\rm eff}^{(2)} - i \gamma_C)^2 - (\delta \gamma + i {\rm{\Delta}})^2} \\
    \delta \gamma &= \gamma_1 - \gamma_2
    \end{split}
\end{equation}

\subsection{Single trajectories considering only local qubit jump operators}

We suppose that a quantum jump $\gamma_1$ or $\gamma_2$ occurs, and the wave function $\ket{\psi (t)}$ is $\ket{\phi} = - i \ket{0, g, e}$ in \eqref{qubit_jump} (the other case being just a relabelling).
For the sake of simplicity, we identify $t$ with the elapsed time after the firt jump took place.

\begin{figure*}
	\centering
	\includegraphics[width=0.9\linewidth]{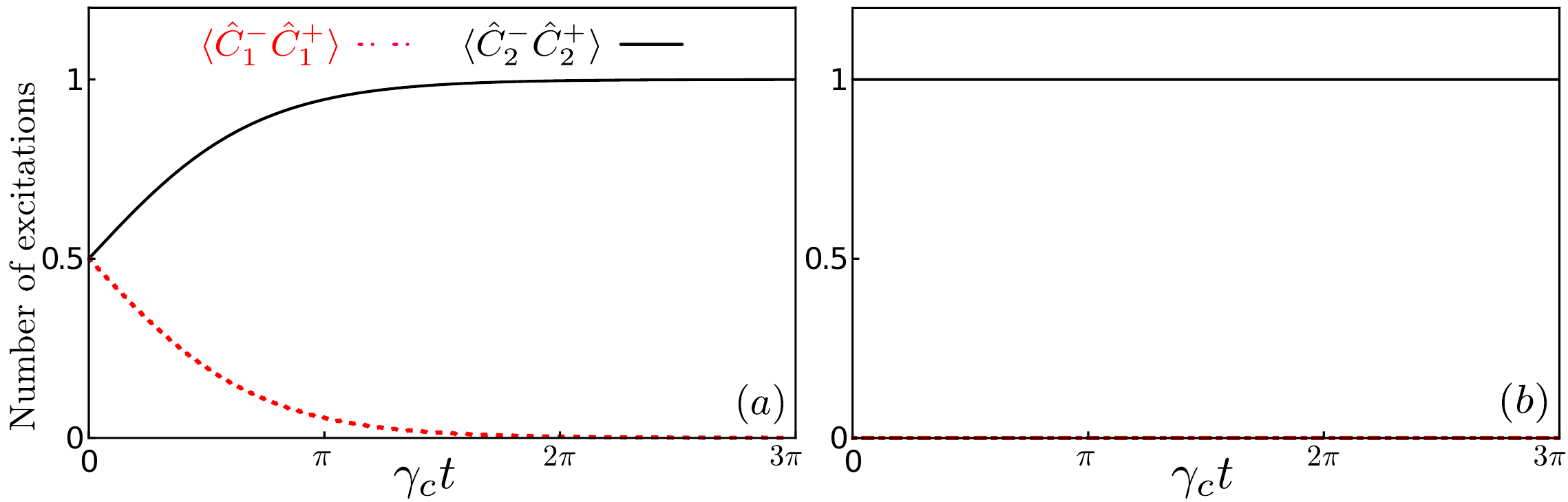}
	\caption{Time evolution of the mean qubit excitation numbers $\expec{\hat C_1^- \hat C_1^+}$ (red dotted curves) and $\expec{\hat C_2^- \hat C_2^+}$ (black solid curves) after:
	(a) A collective qubit jump, as given by Eq.~(\ref{App:Non_oscillate_state_qubit_C1});
	(b) A local qubit 1 jump, as given by Eq.~(\ref{App:Non_oscillate_state_qubit2_C1}). In both cases, the parameters ${2\rm{\Delta}} =\omega_q^{(1)}-\omega_q^{(2)}= 0.3 \omega_0$, $\gamma_1 = \gamma_C = 4 \times 10^{-4} \omega_0$, and $\gamma_2 = 4 \times 10^{-5} \omega_0$ were used.
	\label{App:fig_trajectory4}}
\end{figure*}

The evolution of an initial state $\ket{0, g,e}$ is give by

\be \label{App:second_state} 
\begin{split}
\ket{\phi (t)} &= - i e^{- \frac{1}{4} {\rm{\Gamma}} t} \bigg\{ \mleft[ \cos(\zeta t / 4) + \frac{\delta \gamma}{\zeta} \sin (\zeta t / 4) \mright] \ket{0, g, e}\\
& \quad - \frac{4 i {\rm{\Omega}}_{\rm eff}^{(2)}}{\zeta} \sin (\zeta t / 4) \ket{0, e, g} \bigg \} \, .
\end{split}
\ee

Until another quantum jump occurs, and appropriately renormalizing $\ket{\phi (t)}$, the qubit excitation numbers evolve as 
\be \label{oscillate_state_qubit}
\begin{split}
\expec{\hat C_1^- \hat C_1^+} &= \frac{ \mleft( \frac{4 {\rm{\Omega}}_{\rm eff}^{(2)}}{\zeta} \mright)^2 \sin^2 (\frac{\zeta t }{4})}{1 + \frac{\delta \gamma}{\zeta} \sin (\frac{\zeta t }{ 2}) + 2 \mleft( \frac{\delta \gamma}{\zeta} \mright)^2 \sin^2 (\frac{\zeta t }{4})} \\
\expec{\hat C_2^- \hat C_2^+} &= \frac{\cos^2 (\frac{\zeta t }{ 4}) + \mleft( \frac{\delta \gamma}{\zeta} \mright)^2 \sin^2 (\frac{\zeta t }{4}) + \frac{\delta \gamma}{\zeta} \sin^2 (\frac{\zeta t }{ 2})}{1 + \frac{\delta \gamma}{\zeta} \sin (\frac{\zeta t }{ 2}) + 2 \mleft( \frac{\delta \gamma}{\zeta} \mright)^2 \sin^2 (\frac{\zeta t }{ 4})} \, .
\end{split}
\ee

\subsubsection{Identical qubits}
\label{App:local_identical}
Since $\hat H_{\rm eff}^{(2)}$ in \eqref{Heff_second_oreder} is nonzero for identical qubits, and $\delta \gamma= \Delta=0$, $\ket{0, g, e}$ is not an eigenstate of the system and the oscillations with $\ket{0,e,g}$ are sinusoidal.
In the case shown in \figpanel{fig_trajectory1}{b}, the wave function  $\ket{\phi (t)}$ is projected onto the state $\ket{0, g, g} = |\braket{\phi (t)}{\phi (t)}|^{-1/2}\hat C_i^+ \ket{\phi (t)}$ after $\gamma_1$ emits a second time.

\subsubsection{Non-identical qubits}
\label{App:local_nonidentical}

For non-identical qubits there are two cases to take into consideration. First, if $\Delta=0$ the shape and form of the oscillations depends on the difference between the emission rates $\delta \gamma$. For large values of $\delta \gamma$ the oscillations are completely suppressed. Indeed, the condition $\delta \gamma> 4 {\rm{\Omega}}_{\rm eff}^{(2)}$ makes $\zeta$ imaginary and the oscillations become exponential decays (not shown in the figures).

For ${\rm{\Delta}} \neq 0$, the second-order effective terms $\hat H_{\rm eff}^{(2)}(t)$ in \eqref{Heff_second_oreder} can be neglected thanks to the RWA. The time-evolution operator then acquires the simple form

\be \label{App:evolution_operator3} 
\begin{split}
\hat U(t) &= e^{- \frac{1}{2} \gamma_1 t}  \ketbra{0, e, g}{0, e, g}\\
&+ e^{- \frac{1}{2} \gamma_2 t} \ketbra{0, g, e}{0, g, e} \, .
\end{split}
\ee
Starting from the state $\ket{\phi} = \ket{0, g, e}$ (which now is an eigenstate of the system effective Hamiltonian), the system does not evolve. Thus, the time evolutions of the qubit excitation numbers are simply $\expec{\hat C_1^- \hat C_1^+} = 0$ and $\expec{\hat C_2^- \hat C_2^+} = 1$. This process is shown in \figpanel{fig_trajectory1}{c,d}, where a quantum jump first takes place in qubit 2 (qubit 1) and then in the other qubit. 

\subsection{Single trajectories considering local and collective qubit jump operators}

Here we analyze the case $\gamma_C\neq 0$.
If the first jump is $\gamma_1$, the time-evolution operator in \eqref{evolution_operator2} to the (normalized) initial state $\ket{\phi} = \ket{0, g, e}$ gives

\be \label{App:second_state_collective} 
\begin{split}
 \ket{\phi (t)} &= - i e^{- \frac{1}{4} {\rm{\Gamma}} t}  \bigg \{ \bigg [ \cos (\zeta t / 4) + \frac{\delta \gamma}{\zeta} \sin (\zeta t / 4) \bigg ] \ket{0, g, e} \\
 &\quad - i \frac{4 {\rm{\Omega}}_{\rm eff}^{(2)} - i \gamma_C}{\zeta} \sin (\zeta t / 4) \ket{0, e, g} \bigg \} \, .
 \end{split}
 \ee
If, instead, the first quantum jump is $\gamma_C$, the initial state $\chi^{+}=(\ket{0, g, e} +\ket{0, e,g})/\sqrt{2}$ evolves as
\begin{widetext}
\be \label{second_superposition_state_collective_non_identical2}
\begin{split}
\ket{\chi (t)} =& - \frac{i e^{- \frac{1}{4} \Gamma t}}{\sqrt{2}} \mleft\{ \mleft[ \cos (\zeta t / 4) - i \frac{4 \Omega^{(2)}_{\rm eff} - i\gamma_C}{\zeta} \sin (\zeta t / 4) \mright] \mleft( \ket{0, e, g} + \ket{0, g, e} \mright) \mright. - \mleft. \frac{\delta \gamma + i \Delta}{\zeta} \sin (\zeta t / 4) \mleft( \ket{0, e, g} - \ket{0, g, e} \mright) \mright\} \, ,
\end{split}
\ee
\end{widetext}

We do not report the general formulas for $\expec{\hat C_{1,2}^- \hat C_{1,2}^+}$, but we provide them for the specific cases below

\subsubsection{Identical qubits}\label{App:Collective_identical}

For a $\gamma_1$ jump, and contrary to the case $\gamma_C=0$, this time $\zeta$ is always a complex number, meaning that the system dynamics will have an oscillating part with exponential decay. Considering the case $\gamma =\gamma_1 = \gamma_2 \neq \gamma_C$ as in \figpanel{fig_trajectory3}{a}, \eqref{App:second_state_collective} becomes

\be \label{App:second_state_collective2}
\begin{split}
\ket{\phi (t)} &= \frac{- i e^{- \frac{1}{2} \gamma t} }{2} \bigg \{ e^{ i {\rm{\Omega}}_{\rm eff}^{(2)} t} \bigg [ \ket{0, g, e} - \ket{0, e, g} \bigg ]\\
&\quad + e^{- \frac{1}{2} \gamma_C t} e^{-i {\rm{\Omega}}_{\rm eff}^{(2)} t} \bigg [ \ket{0, g, e} + \ket{0, e, g} \bigg ] \bigg  \} \, .
\end{split}
\ee

Notice that  the symmetric superposition  $\ket{0, g, e} + \ket{0, e, g} $ decays faster than the antisymmetric one due to the factor $e^{- \gamma_C t/2}$. Therefore, the antisymmetric superposition is a dark state of the evolution without quantum jumps, while the symmetric  superposition plays the role of a bright one. With the state $\ket{\phi} = i \ket{0, e, g}$ we end up in the same situation (neglecting a collective phase factor).
Thus, no matter the details of the initial state, normalizing $\ket{\phi (t)}$ in \eqref{App:second_state_collective2}, we see that it tends towards the superposition state
$\ket{\phi (\gamma_C t\gg 1)} \simeq 
( \ket{0, g, e} - \ket{0, e, g})/\sqrt{2}$.

The time evolutions of the qubit excitation numbers are given by
\be \label{App:oscillate_state_qubit_Collective}
\begin{split}
\expec{\hat C_1^- \hat C_1^+} &= \frac{1}{2} - \frac{e^{- \frac{1}{2} \gamma_C t} \mleft[ 1 - 2 \sin^2 \mleft( {\rm{\Omega}}_{\rm eff}^{(2)} t \mright) \mright]}{1 + e^{- \gamma_C t}} \\
\expec{\hat C_2^- \hat C_2^+} &= \frac{1}{2} - \frac{e^{- \frac{1}{2} \gamma_C t} \mleft[ 1 - 2 \cos^2 \mleft( {\rm{\Omega}}_{\rm eff}^{(2)} t \mright) \mright]}{1 + e^{- \gamma_C t}} \, ,
\end{split}
\ee
{\small $$\, $$}
which have sinusoidal oscillations with exponential decay (depending on $\gamma_C$) towards the value $\expec{\hat C_1^- \hat C_1^+} = \expec{\hat C_2^- \hat C_2^+} = 1 / 2$.
The two qubits keep exchanging their excitation around the superposition state $\ket{\phi (\gamma_C t\gg 1)}$ until a collective or local qubit jump occurs, projecting the wave function onto the state $\ket{0, g, g}$, as shown in \figpanel{fig_trajectory3}{a}. 

With the same parameters, the superposition state $\ket{\chi^+}$ resulting from a collective jump [cf. \eqref{collective_qubit_jump}] is an eigenstate of the effective Hamiltonian $\hat H_{\rm eff}$ in \eqref{Heff0}, and $\ket{\chi(t)}$ does not evolve, as shown in \figpanel{fig_trajectory3}{b}.

\begin{figure*}
    \centering
    \includegraphics[width=0.9\linewidth]{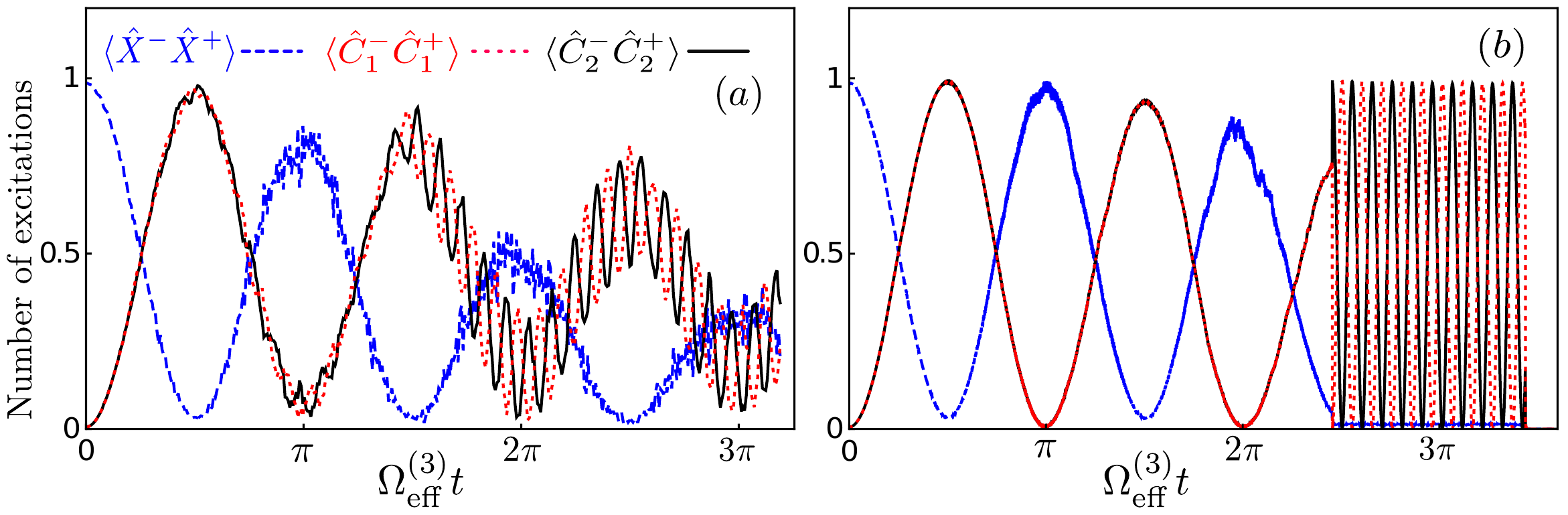}
    \caption{Quantum trajectories for fully and partial homodyne measurement of the system output. The plots show the expectation value of the mean photon number $\expec{\hat X^- \hat X^+}$ (blue dashed curves) and the mean excitation numbers of the two qubits $\expec{\hat C_i^- \hat C_i^+}$ ($i = 1, 2$) (red dotted and black solid curves).
    (a) A quantum trajectory where the output fields of all subsystems are detected through homodyne detection.
    (b) A quantum trajectory where only the output field of the cavity is measured with homodyne detection, while the qubit outputs are measured with photodetection. For both panels, parameters are the same as in \figpanel{fig_trajectory1}{a,b}.
    \label{fig:single_homodyne}}
\end{figure*}

\subsubsection{Non-identical qubits}
\label{App:Collective_nonidentical}

For a local qubit jump $\gamma_1$ and $\ket{\phi} = i \ket{0, g, e}$ the mean qubit excitation numbers are
\begin{widetext}
\be \label{App:Non_oscillate_state_qubit2_C1}
\begin{split}
\expec{\hat C_1^- \hat C_1^+} &= \frac{a'_1 \mleft( \cos[{\rm Im} (\zeta) t / 4] - \cos[{\rm Re} (\zeta) t / 4] \mright)}{c'_1 \cos[{\rm Im} (\zeta) t / 4] + c'_2 \cos[{\rm Re} (\zeta) t / 4] - c'_3 \sin[{\rm Im} (\zeta t / 4] + c'_4 \sin[{\rm Re} (\zeta) t / 4]}  \\
\expec{\hat C_2^- \hat C_2^+} &= \frac{b'_1 \cos[{\rm Im} (\zeta) t / 4] + b'_2 \cos[{\rm Re} (\zeta) t / 4] - c'_3 \sin[{\rm Im} (\zeta) t / 4] + c'_4 \sin[{\rm Re} (\zeta) t / 4]}{c'_1 \cos[{\rm Im} (\zeta) t / 4] + c'_2 \cos[{\rm Re} (\zeta) t / 4] - c'_3 \sin[{\rm Im} (\zeta) t / 4] + c'_4 \sin[{\rm Re} (\zeta) t / 4]} \, ,
\end{split}
\ee
where the coefficients are 
\be \label{coefficients2}
\begin{split}
a'_1 &= \gamma_C^2 \, ,  \\
b'_1 &= \abssq{\zeta} + \delta \gamma^2 + {\rm{\Delta}}^2 \, ,
\quad 
b'_2 = \abssq{\zeta} - \delta \gamma^2 - {\rm{\Delta}}^2 \, ,  \\
c'_1 &= \abssq{\zeta} + \delta \gamma^2 + {\rm{\Delta}}^2 + \gamma_C^2 \, , \quad
c'_2 = \abssq{\zeta} - \delta \gamma^2 - {\rm{\Delta}}^2 - \gamma_C^2 \, , \quad
c'_3 = i {\rm{\Delta}} {\rm Re} (\zeta) + \delta \gamma {\rm Im} (\zeta) \, , \quad
c'_4 = i {\rm{\Delta}} {\rm Im} (\zeta) + \delta \gamma {\rm Re} (\zeta)\, .
\end{split}
\ee
\end{widetext}
In the cases considered in \figpanel{App:fig_trajectory4}{b}, the equations~(\ref{App:Non_oscillate_state_qubit2_C1}) correctlty predict almost no evolution in the system.

When considering instead a collective $\gamma_C$ jump, i.e., the initial state is $\ket{\chi^+}$ in \eqref{second_superposition_state_collective_non_identical}, the mean excitation number of qubits for $\ket{\chi(t)}$ is
\begin{widetext}
\be \label{App:Non_oscillate_state_qubit_C1}
\begin{split}
\expec{\hat C_1^- \hat C_1^+} &= \frac{1}{2} \frac{a_1 \cos[{\rm Im} (\zeta) t / 4] + a_2 \cos[{\rm Re} (\zeta) t / 4] + a_3 \sin[{\rm Im} (\zeta) t / 4] - a_4 \sin[{\rm Re} (\zeta) t / 4]}{c_1 \cos[{\rm Im} (\zeta) t / 4] + c_2 \cos[{\rm Re} (\zeta) t / 4] + c_3 \sin[{\rm Im} (\zeta t / 4] - c_4 \sin[{\rm Re} (\zeta) t / 4]} \\
\expec{\hat C_2^- \hat C_2^+} &= \frac{1}{2} \frac{b_1 \cos[{\rm Im} (\zeta) t / 4] + b_2 \cos[{\rm Re} (\zeta) t / 4] + b_3 \sin[{\rm Im} (\zeta) t / 4] - b_4 \sin[{\rm Re} (\zeta) t / 4]}{c_1 \cos[{\rm Im} (\zeta) t / 4] + c_2 \cos[{\rm Re} (\zeta) t / 4] + c_3 \sin[{\rm Im} (\zeta t / 4] - c_4 \sin[{\rm Re} (\zeta) t / 4]} \, ,
\end{split}
\ee
where the coefficients are 
\be \label{App:coefficients}
\begin{split}
a_1 &= \abssq{\zeta} + (\delta \gamma + {\rm{\Delta}})^2 + \gamma_C^2 \, , \quad
a_2 = \abssq{\zeta} - (\delta \gamma + {\rm{\Delta}})^2 - \gamma_C^2 \, , \\
a_3 &= {\rm Im} (\zeta) (\gamma_C + \delta \gamma) + i {\rm Re} (\zeta) {\rm{\Delta}}\, , \quad
a_4 = {\rm Re} (\zeta) (\gamma_C + \delta \gamma) + i {\rm Im} (\zeta) {\rm{\Delta}}\, ,  
\\
b_1 &= \abssq{\zeta} + (\delta \gamma - {\rm{\Delta}})^2 + \gamma_C^2\, , \quad
b_2 = \abssq{\zeta} - (\delta \gamma - {\rm{\Delta}})^2 - \gamma_C^2 \, , \\
b_3 &= {\rm Im} (\zeta) (\gamma_C - \delta \gamma) - i {\rm Re} (\zeta) {\rm{\Delta}}\, , \quad
b_4 = {\rm Re} (\zeta) (\gamma_C - \delta \gamma) - i {\rm Im} (\zeta) {\rm{\Delta}} \, , \\
c_1 &= \abssq{\zeta} + \delta \gamma^2 + {\rm{\Delta}}^2 + \gamma_C^2 \, , \quad
c_2 = \abssq{\zeta} - \delta \gamma^2 - {\rm{\Delta}}^2 - \gamma_C^2  \, , \quad
c_3 = \gamma_C {\rm Im} (\zeta) \, , \quad
c_4 = \gamma_C {\rm Re} (\zeta) \, .
\end{split}
\ee
\end{widetext}
Equation~(\ref{App:Non_oscillate_state_qubit_C1}) is in agreement in describing a single trajectory after a collective qubit jump has occurred, as shown in  \figpanel{fig_trajectory3}{c,d} for the cases $\gamma_1 > \gamma_2 $ and $\gamma_1 < \gamma_2 $, respectively. For $\gamma_1 = \gamma_2$, $\expec{\hat C_1^- \hat C_1^+} = \expec{\hat C_2^- \hat C_2^+} = 1 / 2$. In this case, the system oscillates between the Bell states  $\ket{0,g,e} \pm \ket{0, e, g}$ and $\ket{0,g,e} \pm i \ket{0, e, g}$ .

\begin{figure*}
	\centering
	\includegraphics[width=0.9\linewidth]{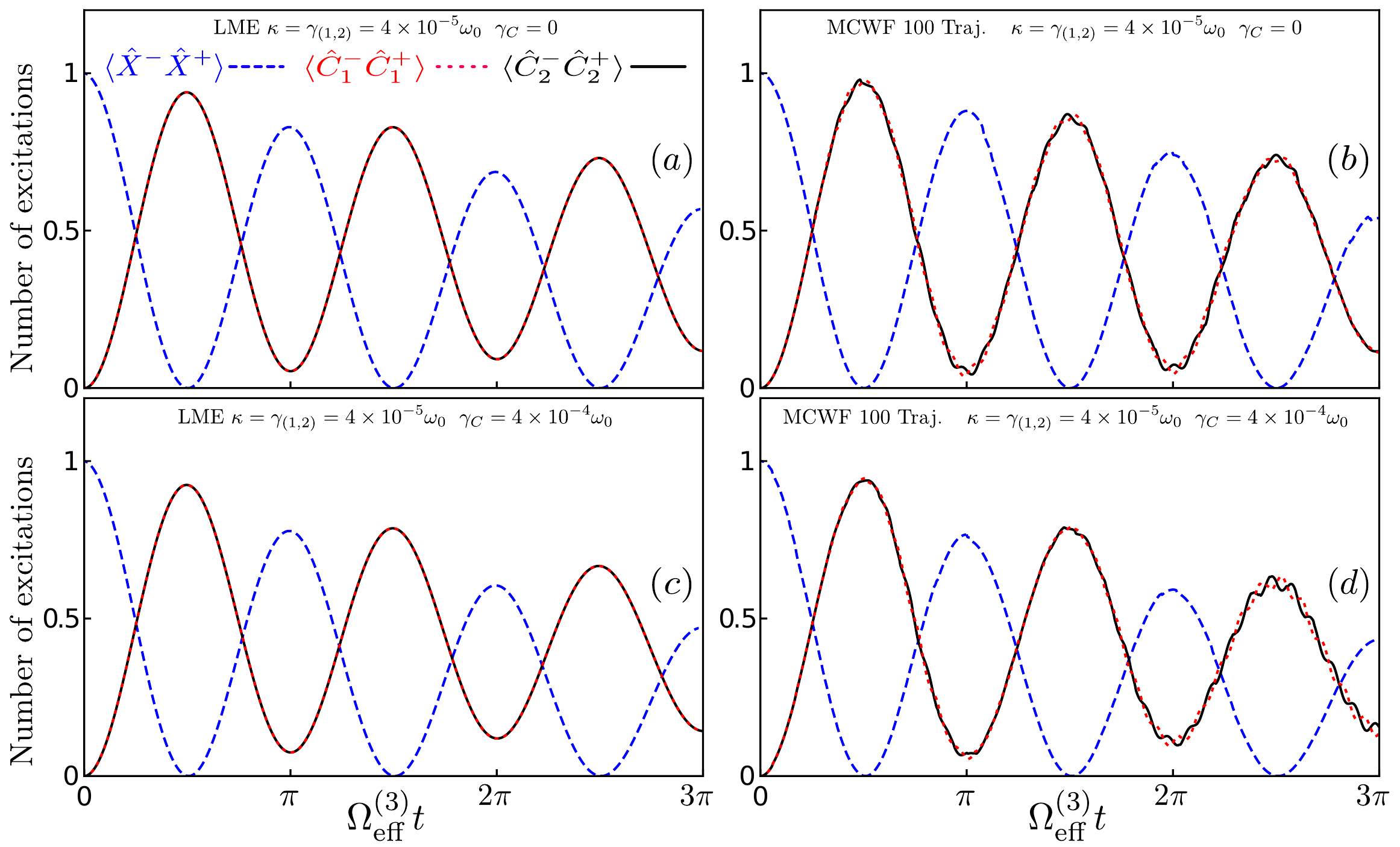}
	\caption{Comparison of system dynamics for the one-photon--two-atom excitation process using the (a,c) LME and (b,d) MCWF approaches. In (a,b) the collective qubit dissipation is $\gamma_C =4 \times 10^{-5} \omega_0$ while in (c,d) $\gamma_C=0$. The plots show the time evolution of the mean photon number $\expec{\hat X^- \hat X^+}$ (blue dashed curves) and the mean excitation numbers of the two qubits $\expec{\hat C_i^- \hat C_i^+}$ ($i = 1, 2$) (red dotted and black solid curves). All the numerical simulations are carried out taking $\ket{1, g, g}$ as the initial state and using the full system Hamiltonian [see \eqref{Hamiltonian} in the main text] near the resonance condition $\omega_c \simeq \omega_q^{(1)} + \omega_q^{(2)}$ for a normalized coupling strength $g = 0.1 \omega_0$.
	\label{App:fig_Dynamics}}
\end{figure*}

\section{Quantum trajectories for homodyne detection}
\label{sec:Homodyne}

To appreciate the importance of the unraveling protocol and of detecting single quantum jumps, let us now consider how the system would evolve under homodyne detection. We can choose to mix a reference coherent field with the output field from either all the subsystems or only some of them. In the continuum limit (infinite amplitude for the reference field), the detectors continuously reads a signal, but the back-action of this signal on the quantum trajectory is minimal. With this protocol, the evolution of the system is diffusive, and dictated by a non-Hermitian Hamiltonian~\cite{Wiseman_BOOK_Quantum}
\be \label{Equation_Homodyne}
\mathcal{\hat{H}_{\rm Hom}} = \mathcal{\hat{H}} - \frac{i}{2} \sum_m \mleft[ \gamma_m^2 \expec{\mleft( \hat{S}_m^{-} - \hat{S}_m^{+} \mright)} + \gamma_m \xi_m (t) \mright] \hat{S}_m^{+} \, ,
\ee
where $\xi_m (t) = dW_m / dt$ is a noise process stemming from the Wiener increment $dW_m$, which has zero mean and variance $dt$. Similarly to quantum trajectories for photodetection, the diffusive stochastic evolution contains the non-Hermitian Hamiltonian $\mathcal{\hat{H}}$ from \eqref{non_Hermitian_H}. However, the effect of quantum jumps is modified by the reference field and enter as the second part of \eqref{Equation_Homodyne}.

Two examples of the resulting diffusive quantum trajectories are plotted in \figref{fig:single_homodyne}, where we re-analyze the one-photon--two-atom excitation process without collective dissipation as plotted in \figpanel{fig_trajectory1}{a,b}. In \figpanel{fig:single_homodyne}{a}, the outputs from all subsystems contribute to the measured homodyne current. In this case, the evolution is damped and no instantaneous change takes place. This demonstrates the importance of the correct unraveling in order to witness all the processes taking place.

To further demonstrate the importance of the collection of the qubit jumps, \figpanel{fig:single_homodyne}{b} shows a trajectory where we detect the cavity output through a homodyne measurement, while the output of the qubits is collected by photodetection. The trajectory shows that a quantum jump of one of the qubits can take place, allowing the two qubits to exchange their remaining excitation as in \figpanel{fig_trajectory1}{b}.

\section{Comparison of system dynamics obtained using the LME and MCWF approaches}
\label{App:comparison_dynamics}

Here, we compare the dynamics of the LME and of averaged MCWF trajectories for the one-photon--two-atom excitation process. In doing this, we consider all the numerical simulations are carried out taking $\ket{1, g, g}$ as the initial state and using the full system Hamiltonian [see \eqref{Hamiltonian} in the main text] near the resonance condition $\omega_c \simeq \omega_q^{(1)} + \omega_q^{(2)}$. In \figpanel{App:fig_Dynamics}{a,b}, we show the main one-photon--two-atom excitation process without collective qubit dissipation included. We clearly see that the MCWF approach (right column) is in complete agreement with the LME approach (left column), which was used in Ref.~\cite{Garziano2016}.
However, the average washes out the qubit-qubit dynamics.
Such a hidden behaviour is completely lost due only to the averaging (the quantum trajectory protocol is identical to the single one shown in the main text).
Since this quantum-jump induced process is fundamental to demonstrate the presence of the main one-photon--two-atom process, it is thus fundamental to collect single trajectories without averaging them.

\bibliography{Riken}

\end{document}